\documentclass[aps,prb,reprint,preprintnumbers]{revtex4-2}
\usepackage[T1]{fontenc}
\usepackage{times}
\usepackage{graphicx,color}
\usepackage{amsfonts,amsmath,amssymb,amsbsy}
\usepackage[colorlinks=true, linkcolor=blue, citecolor=blue, urlcolor=blue]{hyperref}

\def\B{{\mathcal{B}}}
\def\D{{\mathcal{D}}}

\def\I{{\mathcal{I}}}
\let\mathbf=\boldsymbol

\def\blue#1{\textcolor{blue}{#1}}
\def\emph#1{\textcolor{magenta}{#1}}
\begin{document}

\title{Particle-like skyrmions interacting with a funnel obstacle}

\author{Xichao Zhang}
\thanks{These authors contributed equally to this work.}
\affiliation{Department of Electrical and Computer Engineering, Shinshu University, 4-17-1 Wakasato, Nagano 380-8553, Japan}

\author{Jing Xia}
\thanks{These authors contributed equally to this work.}
\affiliation{Department of Electrical and Computer Engineering, Shinshu University, 4-17-1 Wakasato, Nagano 380-8553, Japan}

\author{Xiaoxi Liu}
\email[]{liu@cs.shinshu-u.ac.jp}
\affiliation{Department of Electrical and Computer Engineering, Shinshu University, 4-17-1 Wakasato, Nagano 380-8553, Japan}

\begin{abstract}
The skyrmion-substrate interaction is an important issue to consider when studying particle-like skyrmions in nanostructures, which may strongly affect the skyrmion dynamics and lead to complex dynamic phenomena. Here, we report the current-driven dynamic behaviors of particle-like skyrmions interacting with the funnel obstacle in a ferromagnetic layer. We computationally demonstrate that a funnel obstacle with enhanced perpendicular magnetic anisotropy can be used to modify the dynamics of skyrmions due to the repulsive skyrmion-funnel interaction, which could generate several nontrivial effects, including the compression, clogging, annihilation, guided motion, and capturing of skyrmions. It is also found that the funnel obstacle can separate an incoming flow of skyrmions into two asymmetric flows. Moreover, we show that it is possible to capture skyrmions by the funnel obstacle in both static and dynamic ways. Our results are useful for understanding the skyrmion dynamics interacting with the substrate and could provide guidelines for the control of skyrmion flows in a transmission channel.
\end{abstract}

\date{September 14, 2022}

\preprint{\href{https://doi.org/10.1103/PhysRevB.106.094418}{\textsl{Phys. Rev. B \textbf{106}, 094418 (2022)}}}


\maketitle

\section{Introduction}
\label{se:Introduction}

The particle-substrate interaction is an important issue in the study of active matter systems~\cite{Bechinger_2016,Reichhardt_2017}.
The nanoscale topological skyrmion is a quasiparticle in magnetic systems~\cite{Bogdanov_1989,Roszler_NATURE2006,Lin_PRB2013}.
It can be driven into motion by external forces~\cite{Nagaosa_NNANO2013,Wiesendanger_Review2016,Finocchio_JPD2016,Wanjun_PHYSREP2017,Fert_NATREVMAT2017,Everschor_JAP2018,Zhang_JPCM2020,Gobel_PP2021,Marrows_APL2021,Zhao_2022}, and can also interact with the substrate and show different dynamic behaviors depending on the landscape of substrate~\cite{Reichhardt_2021,Del-Valle_2022}.
For example, in principle a single isolated skyrmion driven by a current in the magnetic layer usually moves in a straight line~\cite{Nagaosa_NNANO2013,Wiesendanger_Review2016,Finocchio_JPD2016,Wanjun_PHYSREP2017,Fert_NATREVMAT2017,Everschor_JAP2018,Zhang_JPCM2020,Gobel_PP2021,Marrows_APL2021,Zhao_2022} and shows the skyrmion Hall effect~\cite{Zang_PRL2011,Wanjun_NPHYS2017,Litzius_NPHYS2017}, when the magnetic layer has no pinning effect and the skyrmion is far away from the layer boundaries.
However, when a pinning landscape is presented in the magnetic layer, the isolated skyrmion may show certain extraordinary dynamic behaviors that are determined by both the pinning and driving forces~\cite{Reichhardt_2021,Del-Valle_2022,Reichhardt_PRL2015,Reichhardt_PRB2015A,Reichhardt_NJP2015,Reichhardt_PRB2015B,Reichhardt_PRB2016,Reichhardt_NJP2016,Reichhardt_PRL2018,Reichhardt_PRB2018,Reichhardt_PRB2020,Vizarim_PRB2020,Muller_PRB2015}.
For example, an isolated skyrmion interacting with asymmetric substrates and ac currents may show Magnus-induced ratchet effects~\cite{Reichhardt_NJP2015}.
It is important to study the skyrmion dynamics in the presence of the skyrmion-substrate interaction as one may use the skyrmion-substrate interaction to control and manipulate the skyrmion dynamics for practical applications~\cite{Reichhardt_2021,Del-Valle_2022}.

Most recently, several experimental reports demonstrated that the perpendicular magnetic anisotropy (PMA) in a magnetic substrate hosting skyrmions could be modified locally in a precise manner~\cite{Juge_NL2021,Ohara_NL2021}, which could be an effective method for constructing an artificial pinning landscape.
The artificial pinning landscape can be used to confine and protect skyrmions~\cite{Juge_NL2021,Ohara_NL2021} and, meanwhile, may affect the static and dynamic properties of skyrmions~\cite{Juge_NL2021,Ohara_NL2021,Zhang_CP2021}.
For example, a computational study suggests that both the position, size, and shape of a nanoscale skyrmion can be modified on a square grid pinning pattern~\cite{Zhang_CP2021}, where the pinning barriers are orthogonal defect lines with reduced PMA.
Therefore, it is envisioned that future skyrmion applications may be based on the manipulation of skyrmions on a well-designed pinning landscape.

Asymmetric obstacles, such as the funnel-shaped pinning barriers, can be used to modify the dynamics of self-driven particles~\cite{Bechinger_2016,Reichhardt_2017}.
The funnel-shaped pinning barriers and obstacles could capture active particles~\cite{Kaiser_PRL2012} and may also lead to the rectification and bidirectional sorting effects in self-driven particle systems~\cite{Wan_PRL2008,Drocco_PRE2012,Berdakin_PRE2013}.
In magnetic systems, funnel obstacles or pinning barriers may also affect the dynamics of particle-like skyrmions driven by external forces.
For example, the dynamics of a single skyrmion or multiple skyrmions can be effectively modified using an array of funnel barriers~\cite{Souza_PRB2021,Souza_2022}, which could lead to the clogging, diode, and collective effects of skyrmions.
However, there are still many open questions in the complex dynamics of multiple skyrmions interacting with funnel obstacles.

In this work, we study the current-induced dynamic behaviors of particle-like nanoscale skyrmions interacting with a funnel obstacle under the framework of micromagnetics.
The funnel obstacle has an enhanced PMA in a ferromagnetic layer with interface-induced Dzyaloshinskii-Moriya interaction (DMI)~\cite{Dzyaloshinsky,Moriya}.
We find that the interactions between the funnel obstacle and the skyrmions are highly nontrivial, which can generate phenomena such as the accumulation, compression, clogging, annihilation, guided motion, and capturing of skyrmions.
The dynamic behaviors of skyrmions depend on the driving force, the incidence angle of skyrmions with respect to the symmetry axis of the funnel obstacle, and the orientation of the funnel obstacle.
Our results suggest that the funnel obstacle could be employed to control the flow of skyrmions in a transmission channel.

\section{Methods}
\label{se:Methods}

We consider a two-dimensional (2D) ferromagnetic layer with interface-induced DMI~\cite{Dzyaloshinsky,Moriya,Tomasello_SREP2014,Xichao_PRB2016B,Sampaio_NN2013} in this work, which can host particle-like N{\'e}el-type skyrmions, i.e., nanoscale and compact skyrmions.
The ferromagnetic layer has a length of $2000$ nm in the $x$ direction and a width of $1000$ nm in the $y$ direction. The thickness of the ferromagnetic layer is fixed at $1$ nm.
The periodic boundary condition is applied in the $y$ direction of the ferromagnetic layer, while the open boundary condition is applied in the $x$ direction.
We assume that a heavy metal layer is underneath the ferromagnetic layer, which generates the interfacial DMI and is also responsible for the generation of the driving force due to the spin Hall effect~\cite{Tomasello_SREP2014,Sampaio_NN2013,Wanjun_SCIENCE2015,Xichao_PRB2016B}.
Thus the magnetization dynamics in the ferromagnetic layer is controlled by the Landau-Lifshitz-Gilbert (LLG) equation augmented with the damping-like spin-orbit torque~\cite{Sampaio_NN2013,Tomasello_SREP2014,Wanjun_SCIENCE2015,Xichao_PRB2016B},
\begin{equation}
\label{eq:LLGS-CPP}
\partial_{t}\boldsymbol{m}=-\gamma_{0}\boldsymbol{m}\times\boldsymbol{h}_{\text{eff}}+\alpha(\boldsymbol{m}\times\partial_{t}\boldsymbol{m})+\boldsymbol{\tau}_{\text{d}},
\end{equation}
where $\boldsymbol{m}$ is the reduced magnetization,
$t$ is the time,
$\gamma_0$ is the absolute gyromagnetic ratio,
$\alpha$ is the Gilbert damping parameter, and 
$\boldsymbol{h}_{\rm{eff}}=-\frac{1}{\mu_{0}M_{\text{S}}}\cdot\frac{\delta\varepsilon}{\delta\boldsymbol{m}}$ is the effective field with $\mu_{0}$, $M_{\text{S}}$, and $\varepsilon$ being the vacuum permeability constant, saturation magnetization, and average energy density, respectively.
The energy terms in our model include the ferromagnetic exchange energy, interfacial DMI energy, PMA energy, and demagnetization energy~\cite{Sampaio_NN2013,Tomasello_SREP2014,Xichao_PRB2016B}.

For the sake of simplicity, we consider the damping-like spin-orbit torque as the driving force, which is expressed as
$\boldsymbol{\tau}_{\text{d}}=u\left(\boldsymbol{m}\times\boldsymbol{p}\times\boldsymbol{m}\right)$ with the coefficient being $u=\left|\left(\gamma_{0}\hbar/\mu_{0}e\right)\right|\cdot\left(j\theta_{\text{SH}}/2aM_{\text{S}}\right)$, where
$\hbar$ is the reduced Planck constant, $e$ is the electron charge, $a$ is the sample thickness, $j$ is the current density, and $\theta_{\text{SH}}$ is the spin Hall angle.
We note that we do not include the field-like spin-orbit torque in Eq.~(\ref{eq:LLGS-CPP}) as we only focus on the dynamics of compact and rigid skyrmions driven by a small or moderate force in this work~\cite{Tomasello_SREP2014}. The field-like torque usually results in the deformation of a skyrmion at a large driving force~\cite{Litzius_NPHYS2017}.

The magnetic parameters are adopted from Refs.~\onlinecite{Sampaio_NN2013,Tomasello_SREP2014,Xichao_PRB2016B}: $\gamma_{0}=2.211\times 10^{5}$ m A$^{-1}$ s$^{-1}$, $\alpha=0.3$, $M_{\text{S}}=580$ kA m$^{-1}$, the exchange constant $A=15$ pJ m$^{-1}$, the PMA constant $K=0.8$ MJ m$^{-3}$, and the DMI constant $D=3$ mJ m$^{-2}$.
We assume that $\theta_{\text{SH}}=1$ so that the driving force is simply determined by the current density $j$ (i.e., $u\sim j$).
The simulations are performed by the \textsc{mumax$^3$} micromagnetic simulator~\cite{MuMax} on a commercial graphics processing unit NVIDIA GeForce RTX 3080 using the driver of version 512.15.
The mesh size in all simulations is $2.5$ $\times$ $2.5$ $\times$ $1$ nm$^3$, which ensures good computational accuracy and efficiency.

\begin{figure}[t]
\centerline{\includegraphics[width=0.48\textwidth]{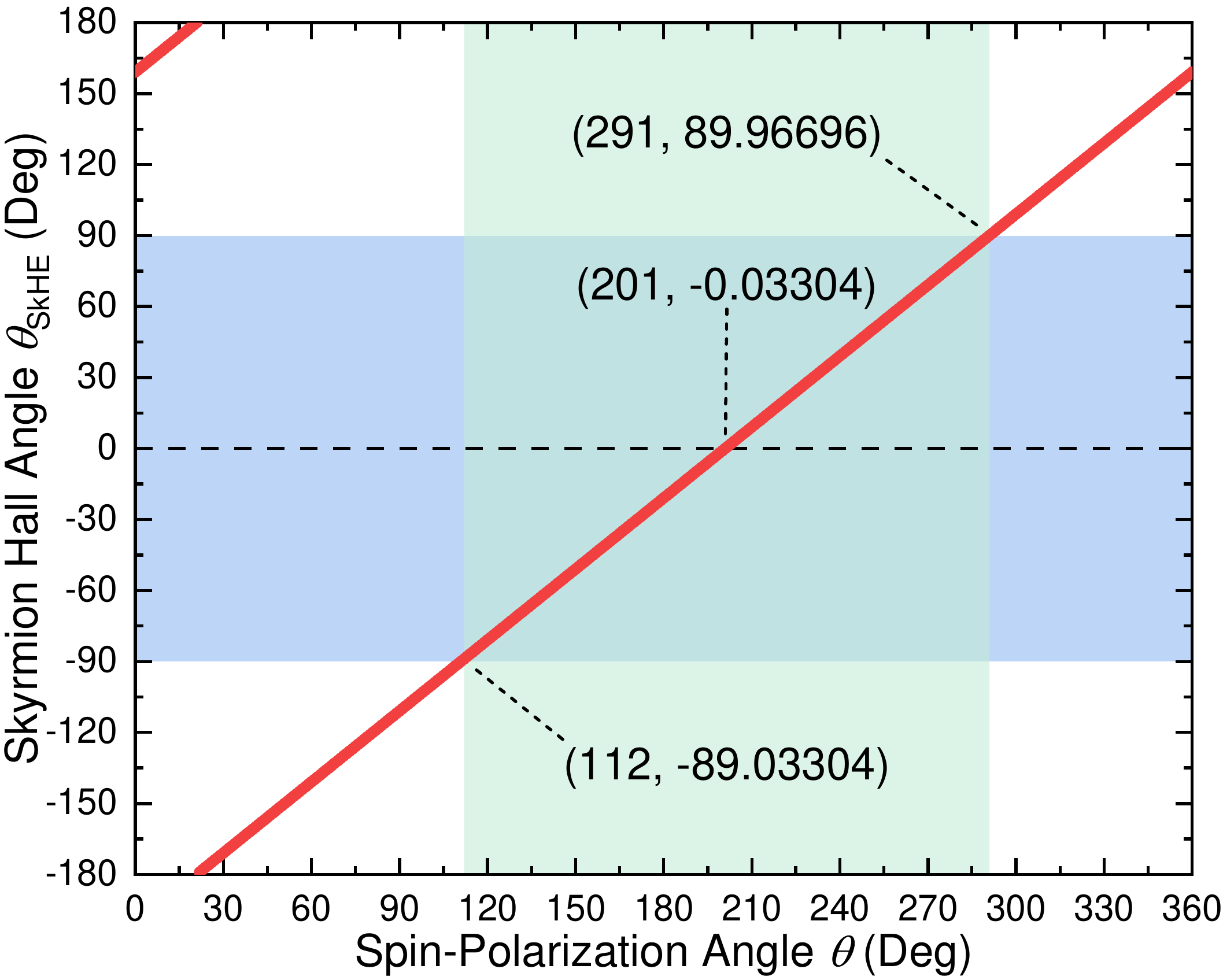}}
\caption{%
Skyrmion Hall angle $\theta_{\text{SkHE}}$ of a single isolated skyrmion as a function of the spin-polarization angle $\theta$. We focus on the skyrmion motion toward the right side of the ferromagnetic layer in order to study the interaction between the skyrmions and the funnel obstacle, i.e., $\theta_{\text{SkHE}}=-90^{\circ}-90^{\circ}$ (indicated in blue). The corresponding range of the spin-polarization angle is $\theta=112^{\circ}-291^{\circ}$ (indicated in green). When $\theta=201^{\circ}$, the skyrmion moves toward the $+x$ direction, i.e., $\theta_{\text{SkHE}}\sim 0^{\circ}$.
}
\label{FIG1}
\end{figure}

\begin{figure*}[t]
\centerline{\includegraphics[width=0.98\textwidth]{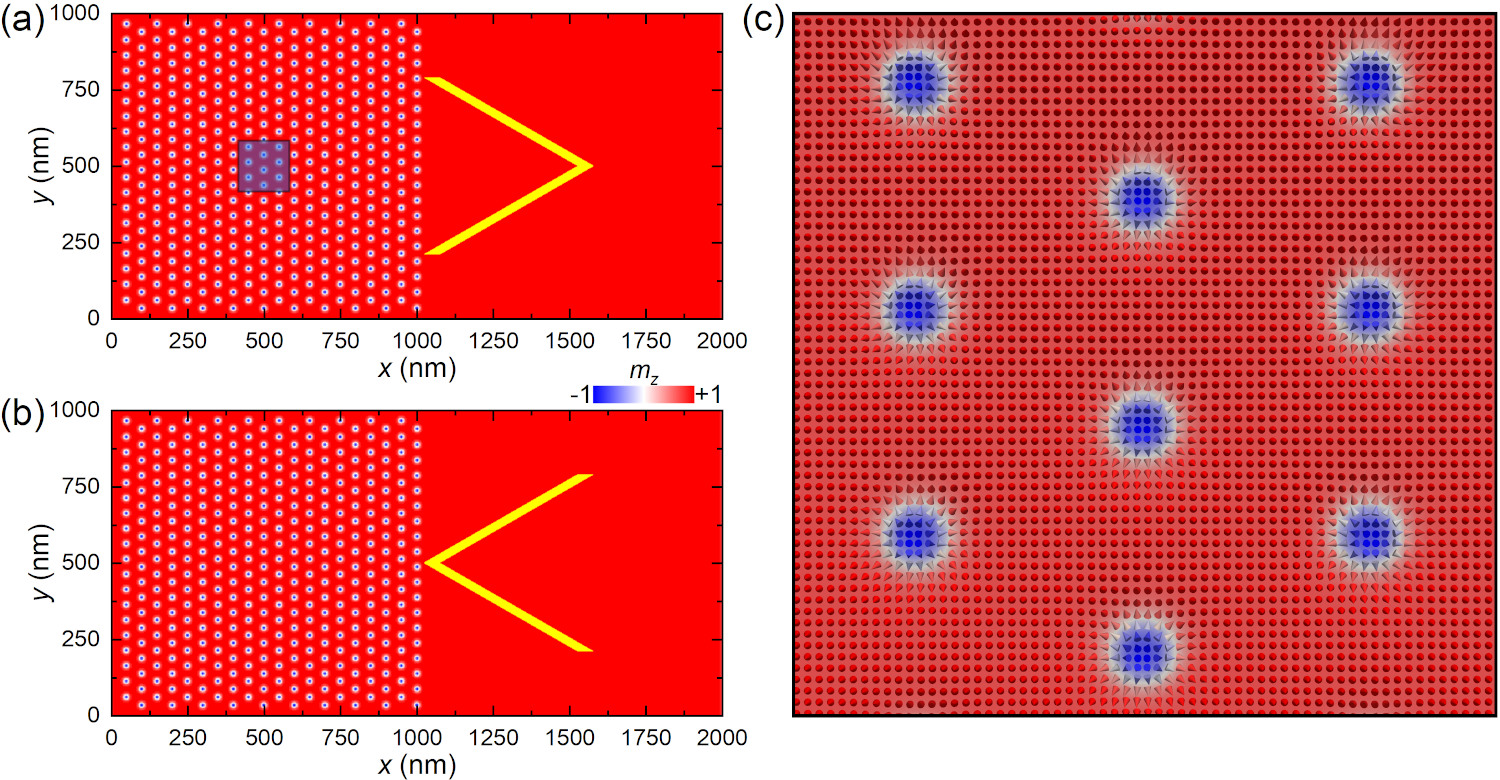}}
\caption{%
Simulation model and initial spin configurations.
(a) Top view of the ferromagnetic layer with a funnel obstacle with its apex pointing to the $+x$ direction (indicated in yellow). The funnel obstacle is placed at the right side of the ferromagnetic layer, which has an apex angle of $60^{\circ}$ and a tip-to-tip width of $\sim 578$ nm. The initial spin configuration at $t=0$ ps is a triangular lattice of $380$ relaxed particle-like skyrmions with $Q=-1$ at the left side of the ferromagnetic layer.
The length, width, and thickness of the ferromagnetic layer equal $2000$ nm, $1000$ nm, and $1$ nm, respectively. The driving force will be applied to the whole ferromagnetic layer. We focus on the skyrmion dynamics around the funnel obstacle.
The color scale represents the reduced out-of-plane spin component $m_z$, which has been used throughout the work.
(b) Top view of the ferromagnetic layer with a funnel obstacle with its apex pointing to the $-x$ direction (indicated in yellow). The parameters of the funnel obstacle are the same as that in (a), except the orientation.
(c) Close-up top view of the skyrmions, corresponding to the region highlighted by a blue box in (a). Each cone represents a spin.
}
\label{FIG2}
\end{figure*}

\begin{figure*}[t]
\centerline{\includegraphics[width=0.98\textwidth]{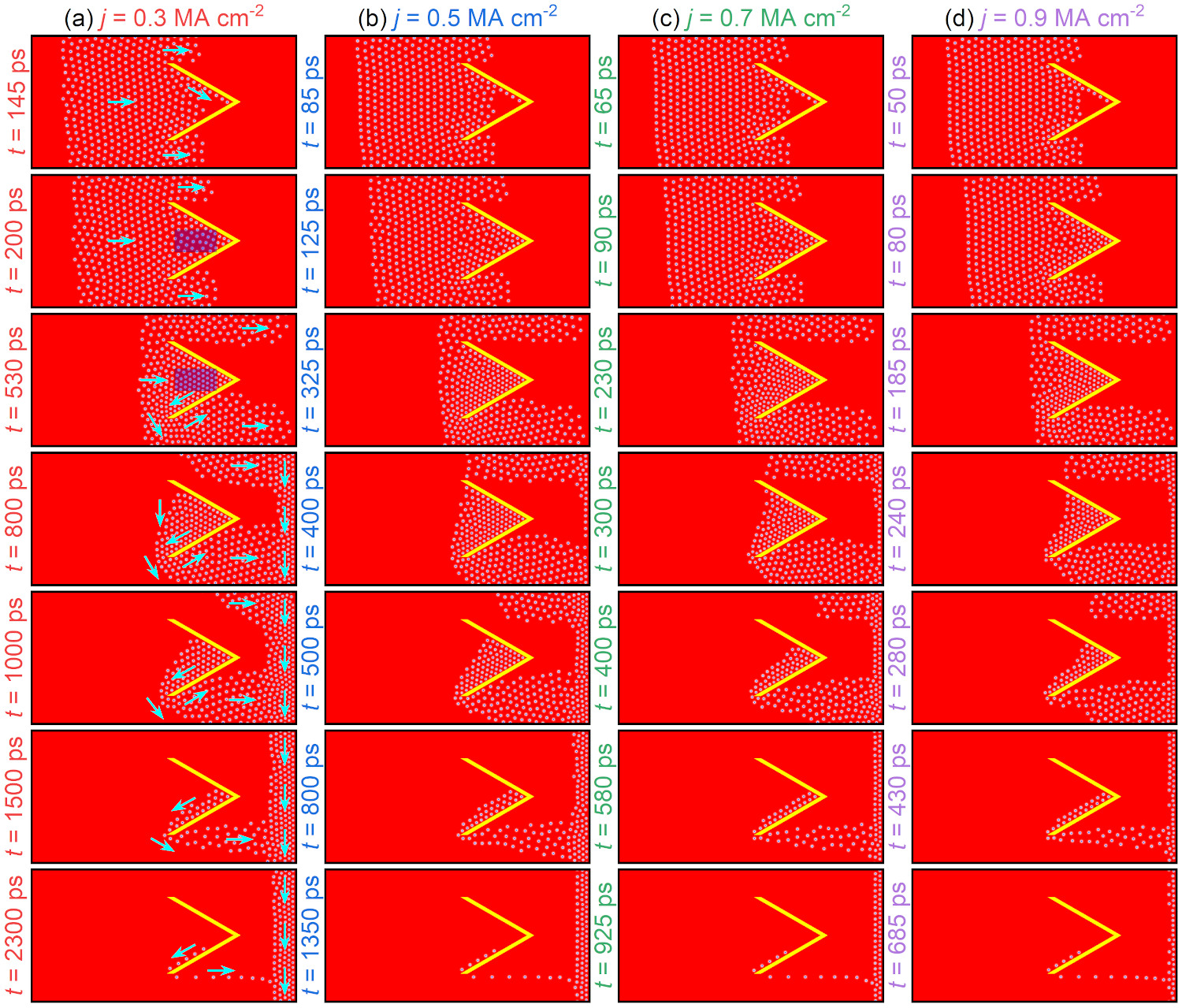}}
\caption{%
Selected top-view snapshots for the systems driven by different current densities $j$ with the spin-polarization angle $\theta=201^{\circ}$. The funnel obstacle is indicated in yellow. The apex of the funnel obstacle is pointing to the $+x$ direction. The intrinsic skyrmion Hall angle $\theta_{\text{SkHE}}=0^{\circ}$.
(a) A current of $j=0.3$ MA cm$^{-2}$ is applied to drive skyrmions. Top-view snapshots at $t=145$, $200$, $530$, $800$, $1000$, $1500$, and $2300$ ps are given.
(b) A current of $j=0.5$ MA cm$^{-2}$ is applied to drive skyrmions. Top-view snapshots at $t=85$, $125$, $325$, $400$, $500$, $800$, and $1350$ ps are given.
(c) A current of $j=0.7$ MA cm$^{-2}$ is applied to drive skyrmions. Top-view snapshots at $t=65$, $90$, $230$, $300$, $400$, $580$, and $925$ ps are given.
(d) A current of $j=0.9$ MA cm$^{-2}$ is applied to drive skyrmions. Top-view snapshots at $t=50$, $80$, $185$, $240$, $280$, $430$, and $685$ ps are given.
The cyan arrow in (a) indicates the motion direction of skyrmions. The motion direction of skyrmions in (b)-(d) is similar to that in (a).
}
\label{FIG3}
\end{figure*}

\begin{figure*}[t]
\centerline{\includegraphics[width=0.98\textwidth]{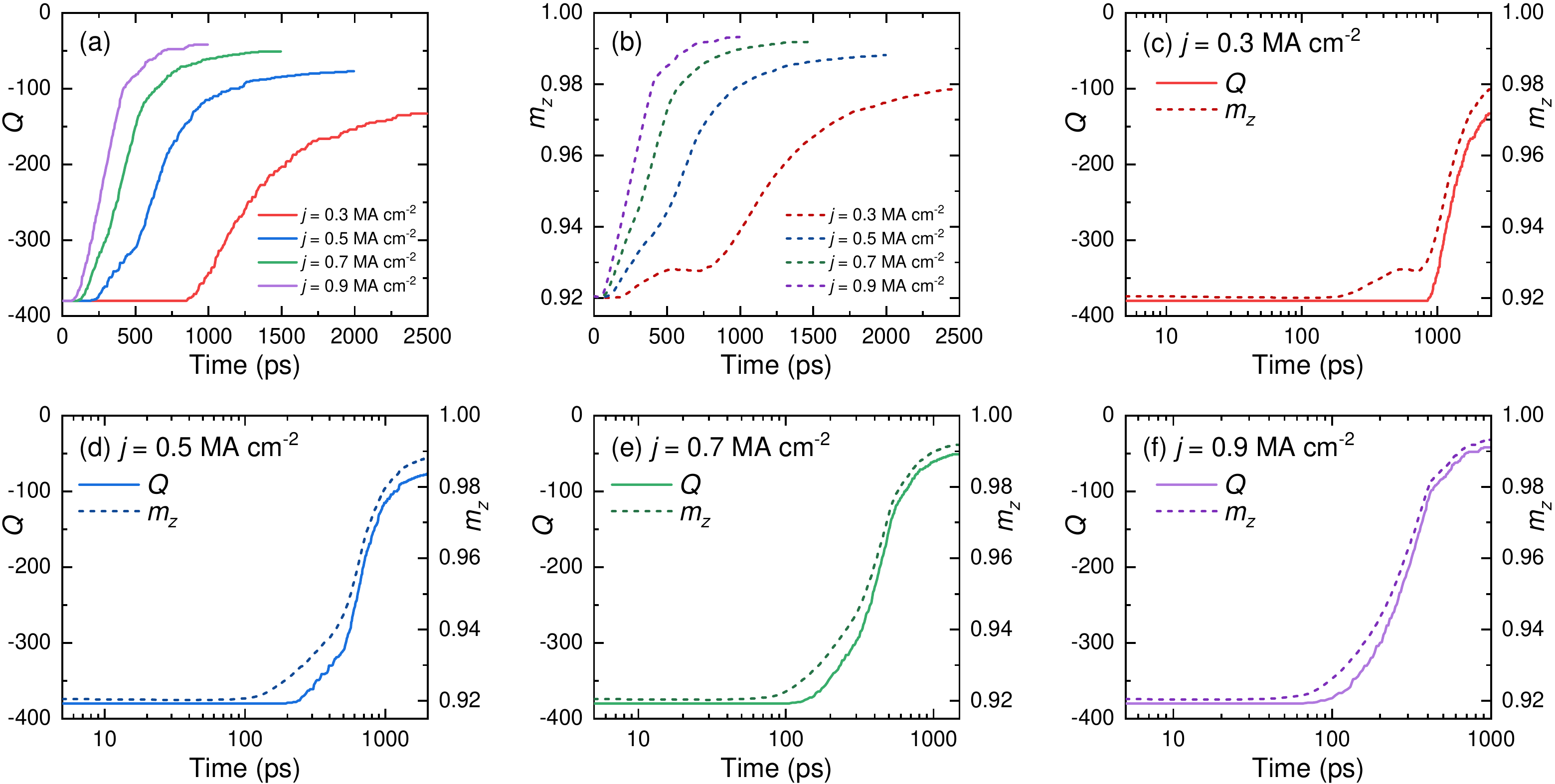}}
\caption{%
Time-dependent skyrmion number $Q$ and reduced out-of-plane magnetization $m_z$ for the systems driven by different current densities $j$ with the spin-polarization angle $\theta=201^{\circ}$. The apex of the funnel obstacle is pointing to the $+x$ direction. The intrinsic skyrmion Hall angle $\theta_{\text{SkHE}}=0^{\circ}$.
(a) Time-dependent $Q$ for different $j$.
(b) Time-dependent $m_z$ for different $j$.
(c) Time-dependent $Q$ and $m_z$ for $j=0.3$ MA cm$^{-2}$.
(d) Time-dependent $Q$ and $m_z$ for $j=0.5$ MA cm$^{-2}$.
(e) Time-dependent $Q$ and $m_z$ for $j=0.7$ MA cm$^{-2}$.
(f) Time-dependent $Q$ and $m_z$ for $j=0.9$ MA cm$^{-2}$.
The simulation time is $2500$, $2000$, $1500$, and $1000$ ps, for the systems driven by $j=0.3$, $0.5$, $0.7$, and $0.9$ MA cm$^{-2}$, respectively.
}
\label{FIG4}
\end{figure*}

\begin{figure}[t]
\centerline{\includegraphics[width=0.48\textwidth]{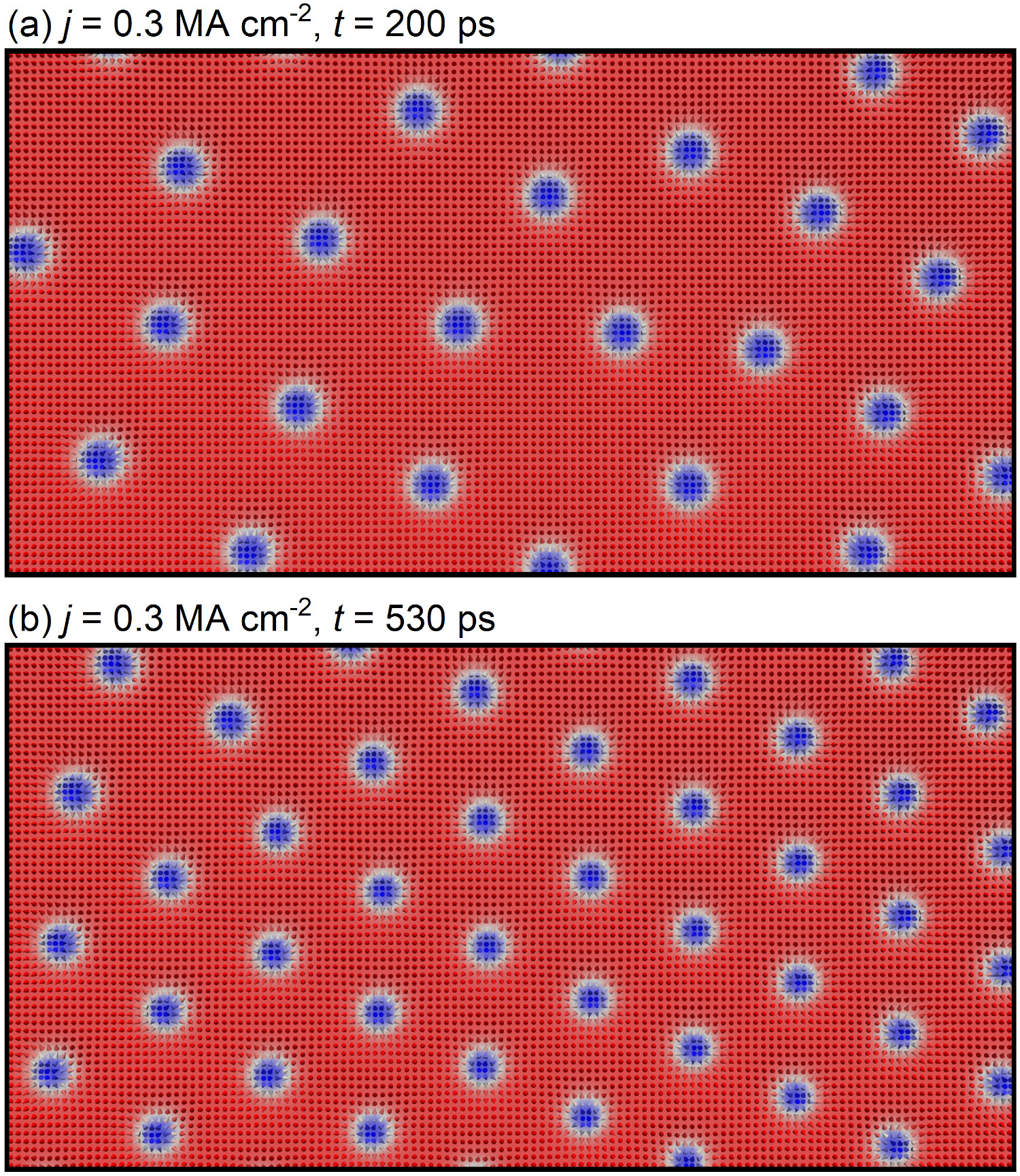}}
\caption{%
Close-up top view of the spin configuration of skyrmions inside the triangular region between the upper and lower funnel walls, corresponding to the region highlighted by a blue box in Fig.~\ref{FIG4}(a), where $j=0.3$ MA cm$^{-2}$ and $\theta=201^{\circ}$. Each cone represents a spin.
(a) The spin configuration of skyrmions at $t=200$ ps.
(b) The spin configuration of skyrmions at $t=530$ ps.
The skyrmion-funnel interaction leads to a compression effect inside the triangular region between the upper and lower funnel walls, where a compressed triangular lattice of skyrmions is formed.
}
\label{FIG5}
\end{figure}

\begin{figure*}[t]
\centerline{\includegraphics[width=0.98\textwidth]{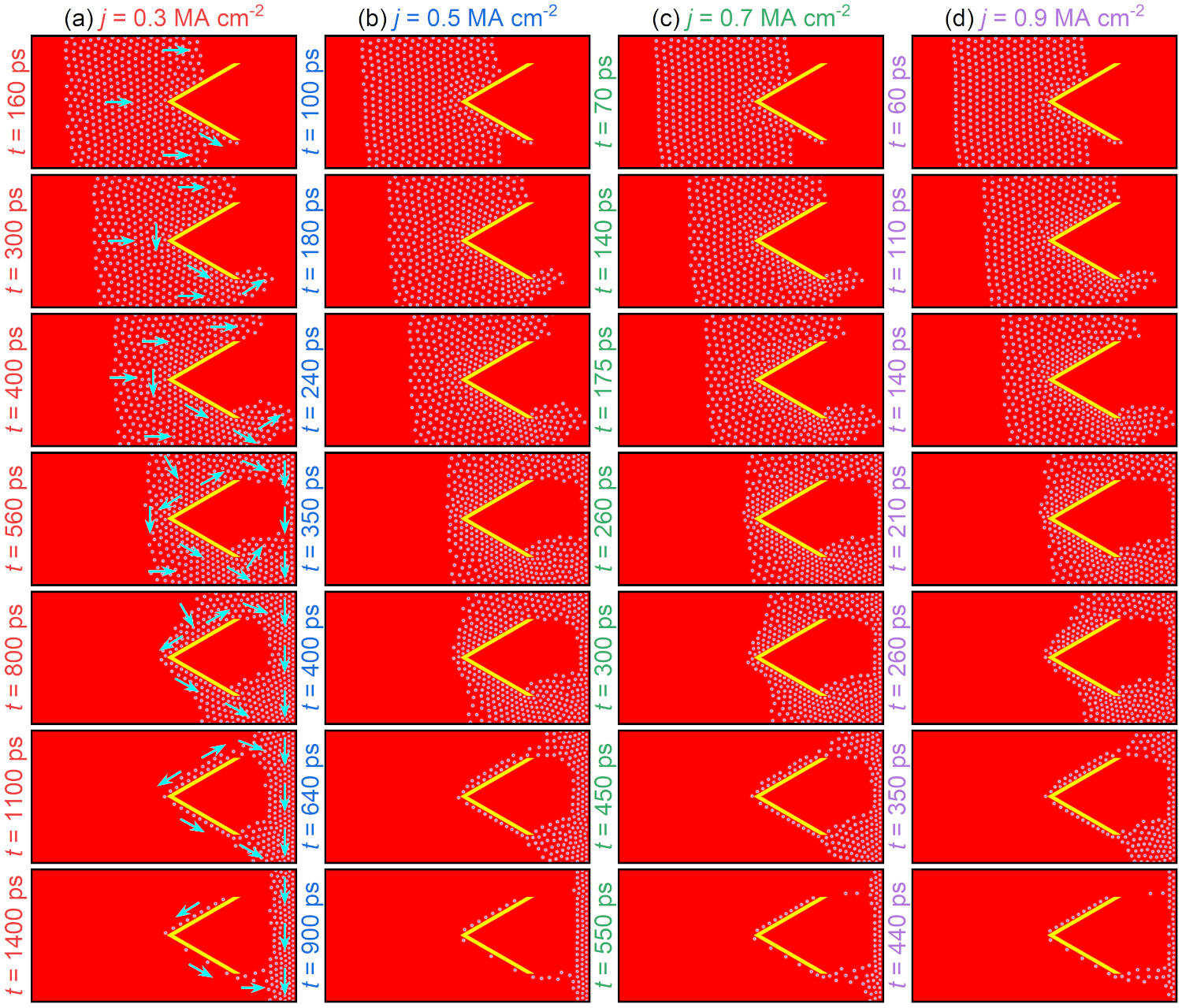}}
\caption{%
Selected top-view snapshots for the systems driven by different current densities $j$ with the spin-polarization angle $\theta=201^{\circ}$. The funnel obstacle is indicated in yellow. The apex of the funnel obstacle is pointing to the $-x$ direction. The intrinsic skyrmion Hall angle $\theta_{\text{SkHE}}=0^{\circ}$.
(a) A current of $j=0.3$ MA cm$^{-2}$ is applied to drive skyrmions. Top-view snapshots at $t=160$, $300$, $400$, $560$, $800$, $1100$, and $1400$ ps are given.
(b) A current of $j=0.5$ MA cm$^{-2}$ is applied to drive skyrmions. Top-view snapshots at $t=100$, $180$, $240$, $350$, $400$, $640$, and $900$ ps are given.
(c) A current of $j=0.7$ MA cm$^{-2}$ is applied to drive skyrmions. Top-view snapshots at $t=70$, $140$, $175$, $260$, $300$, $450$, and $550$ ps are given.
(d) A current of $j=0.9$ MA cm$^{-2}$ is applied to drive skyrmions. Top-view snapshots at $t=60$, $110$, $140$, $210$, $260$, $350$, and $440$ ps are given.
The cyan arrow in (a) indicates the motion direction of skyrmions. The motion direction of skyrmions in (b)-(d) is similar to that in (a).
}
\label{FIG6}
\end{figure*}

\begin{figure*}[t]
\centerline{\includegraphics[width=0.98\textwidth]{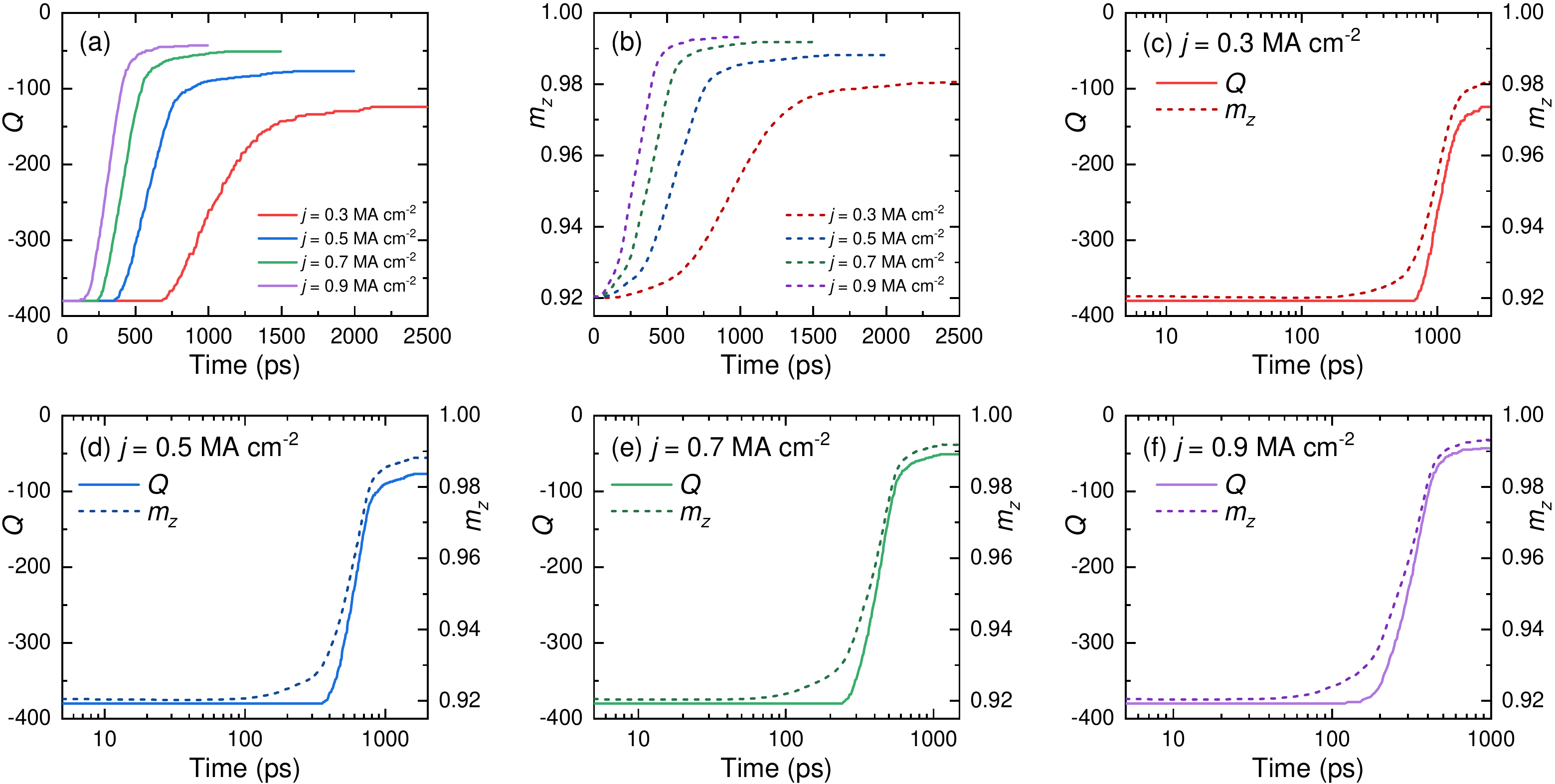}}
\caption{%
Time-dependent skyrmion number $Q$ and reduced out-of-plane magnetization $m_z$ for the systems driven by different current densities $j$ with the spin-polarization angle $\theta=201^{\circ}$. The apex of the funnel obstacle is pointing to the $-x$ direction. The intrinsic skyrmion Hall angle $\theta_{\text{SkHE}}=0^{\circ}$.
(a) Time-dependent $Q$ for different $j$.
(b) Time-dependent $m_z$ for different $j$.
(c) Time-dependent $Q$ and $m_z$ for $j=0.3$ MA cm$^{-2}$.
(d) Time-dependent $Q$ and $m_z$ for $j=0.5$ MA cm$^{-2}$.
(e) Time-dependent $Q$ and $m_z$ for $j=0.7$ MA cm$^{-2}$.
(f) Time-dependent $Q$ and $m_z$ for $j=0.9$ MA cm$^{-2}$.
The simulation time is $2500$, $2000$, $1500$, and $1000$ ps, for the systems driven by $j=0.3$, $0.5$, $0.7$, and $0.9$ MA cm$^{-2}$, respectively.
}
\label{FIG7}
\end{figure*}

\begin{figure*}[t]
\centerline{\includegraphics[width=0.98\textwidth]{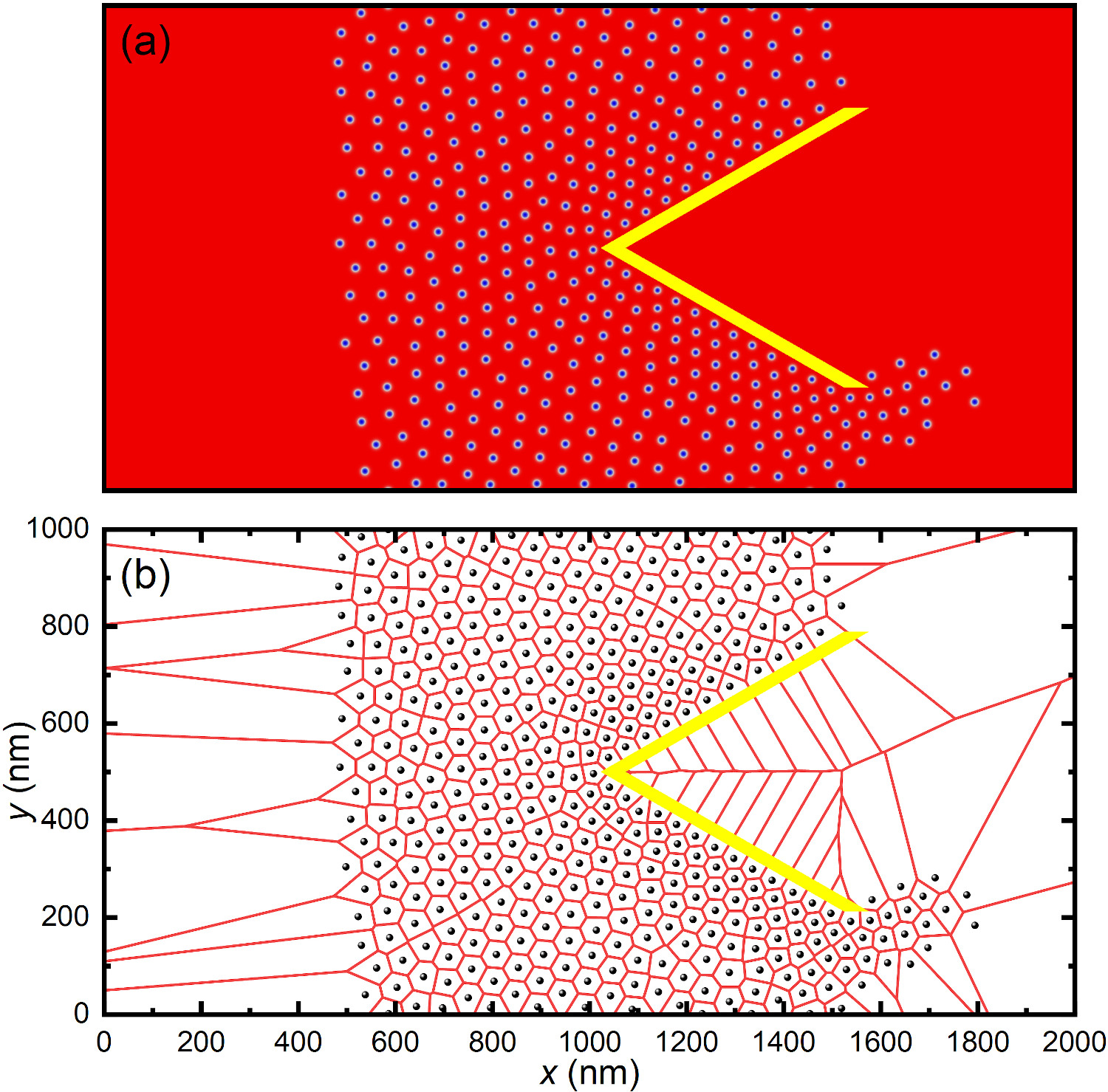}}
\caption{%
(a) Enlarged top-view snapshot for the system driven by $j=0.5$ MA cm$^{-2}$ at $t=180$ ps [see Fig.~\ref{FIG6}(b)]. The funnel obstacle is indicated in yellow. The apex of the funnel obstacle is pointing to the $-x$ direction.
(b) Voronoi cell construction showing the location of the skyrmions in (a). Skyrmion positions are indicated by black dots.
The skyrmions near the funnel walls form triangular lattices. Besides, the density of skyrmions near the funnel walls is higher than that of the incoming skyrmions at the left side of the sample due to the compression effect.
}
\label{FIG8}
\end{figure*}

\begin{figure*}[t]
\centerline{\includegraphics[width=0.98\textwidth]{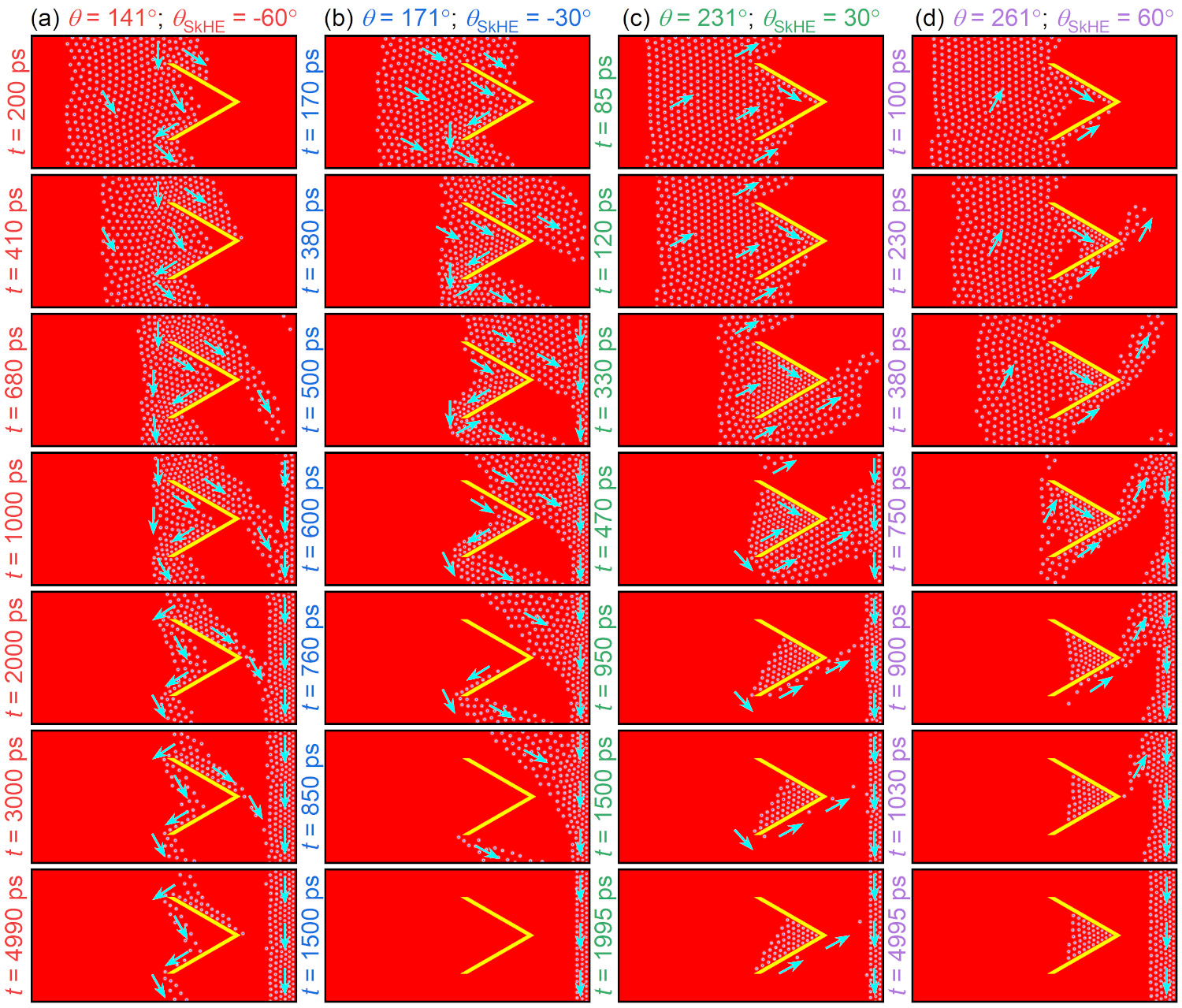}}
\caption{%
Selected top-view snapshots for the systems driven by $j=0.5$ MA cm$^{-2}$ with different spin-polarization angle $\theta$. The funnel obstacle is indicated in yellow. The apex of the funnel obstacle is pointing to the $+x$ direction. The cyan arrow indicates the direction of motion of skyrmions.
(a) A current with $\theta=141^{\circ}$ is applied to drive skyrmions. The intrinsic skyrmion Hall angle $\theta_{\text{SkHE}}=-60^{\circ}$. Top-view snapshots at $t=200$, $410$, $680$, $1000$, $2000$, $3000$, and $4990$ ps are given.
(b) A current with $\theta=171^{\circ}$ is applied to drive skyrmions. The intrinsic skyrmion Hall angle $\theta_{\text{SkHE}}=-30^{\circ}$. Top-view snapshots at $t=170$, $380$, $500$, $600$, $760$, $850$, and $1500$ ps are given.
(c) A current with $\theta=231^{\circ}$ is applied to drive skyrmions. The intrinsic skyrmion Hall angle $\theta_{\text{SkHE}}=30^{\circ}$. Top-view snapshots at $t=85$, $120$, $330$, $470$, $950$, $1500$, and $1995$ ps are given.
(d) A current with $\theta=261^{\circ}$ is applied to drive skyrmions. The intrinsic skyrmion Hall angle $\theta_{\text{SkHE}}=60^{\circ}$. Top-view snapshots at $t=100$, $230$, $380$, $750$, $900$, $1030$, and $4995$ ps are given.
}
\label{FIG9}
\end{figure*}

\begin{figure*}[t]
\centerline{\includegraphics[width=0.98\textwidth]{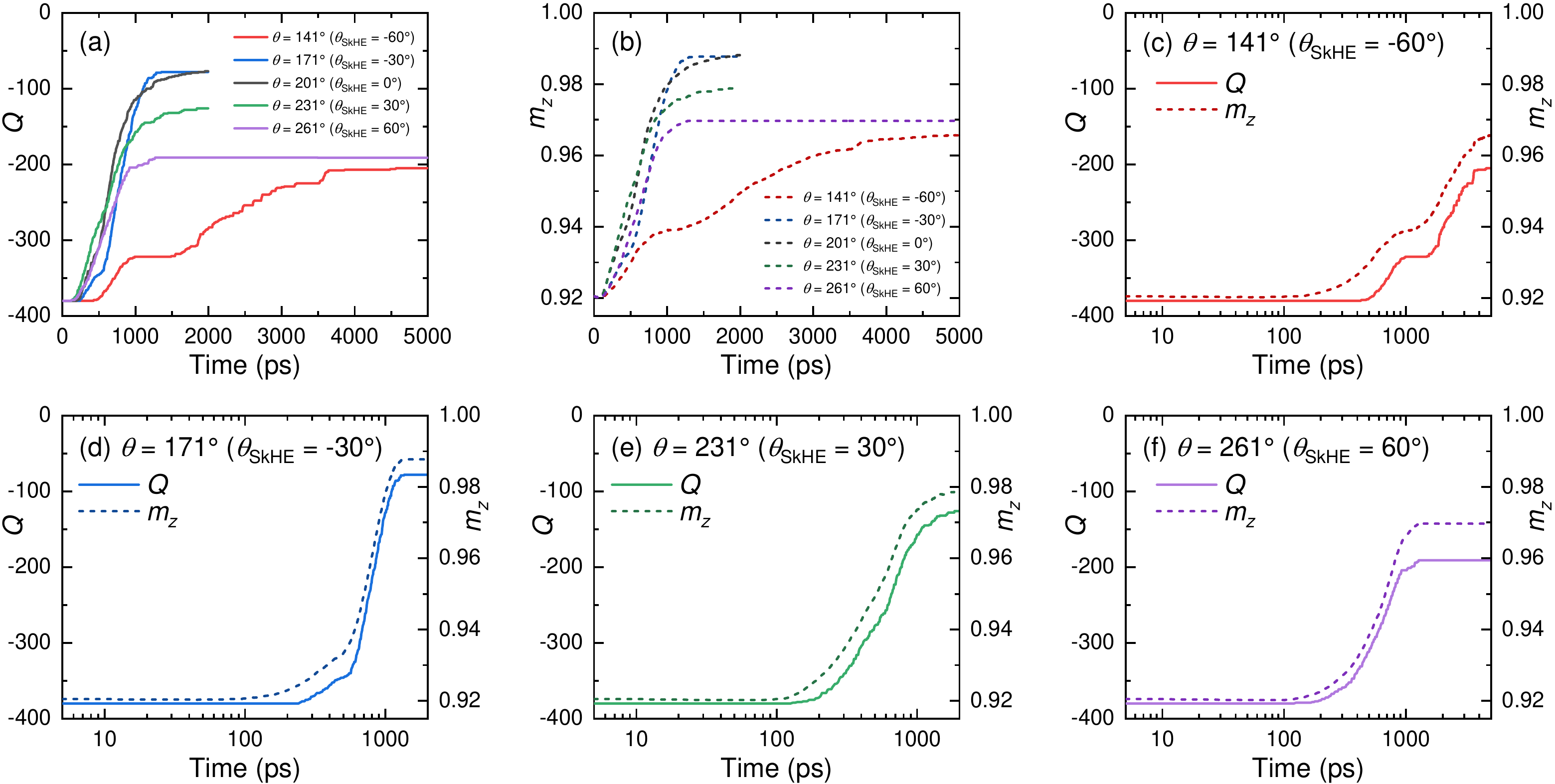}}
\caption{%
Time-dependent skyrmion number $Q$ and reduced out-of-plane magnetization $m_z$ for the systems driven by $j=0.5$ MA cm$^{-2}$ with different spin-polarization angle $\theta$. The apex of the funnel obstacle is pointing to the $+x$ direction.
(a) Time-dependent $Q$ for different $\theta$.
(b) Time-dependent $m_z$ for different $\theta$.
(c) Time-dependent $Q$ and $m_z$ for $\theta=141^{\circ}$, where the intrinsic skyrmion Hall angle $\theta_{\text{SkHE}}=-60^{\circ}$.
(d) Time-dependent $Q$ and $m_z$ for $\theta=171^{\circ}$, where $\theta_{\text{SkHE}}=-30^{\circ}$.
(e) Time-dependent $Q$ and $m_z$ for $\theta=231^{\circ}$, where $\theta_{\text{SkHE}}=30^{\circ}$.
(f) Time-dependent $Q$ and $m_z$ for $\theta=261^{\circ}$, where $\theta_{\text{SkHE}}=60^{\circ}$.
The simulation time is $5000$, $2000$, $2000$, and $5000$ ps, for the systems with $\theta=141^{\circ}$, $171^{\circ}$, $231^{\circ}$, and $261^{\circ}$, respectively.
}
\label{FIG10}
\end{figure*}

\begin{figure*}[t]
\centerline{\includegraphics[width=0.98\textwidth]{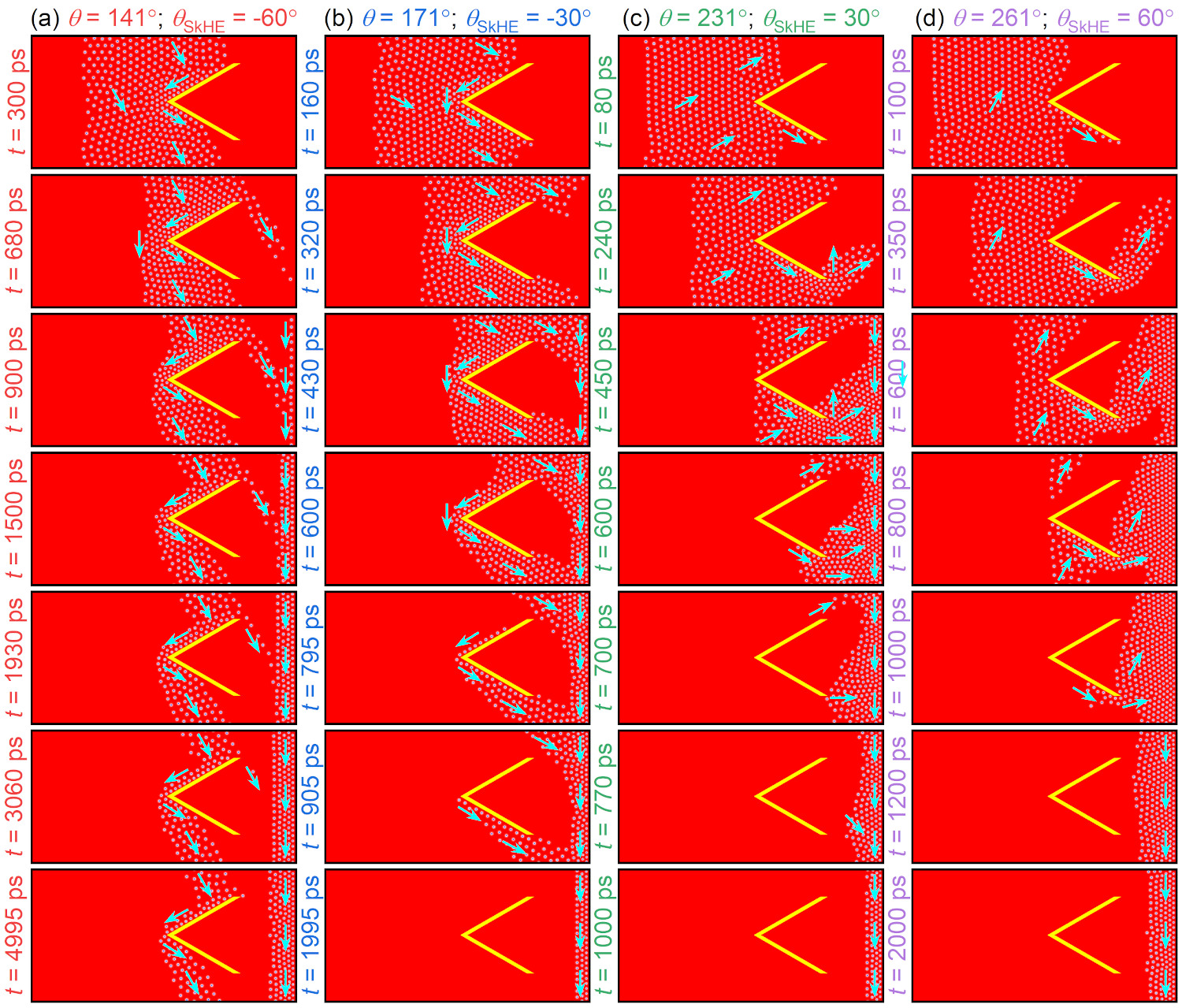}}
\caption{%
Selected top-view snapshots for the systems driven by $j=0.5$ MA cm$^{-2}$ with different spin-polarization angle $\theta$. The funnel obstacle is indicated in yellow. The apex of the funnel obstacle is pointing to the $-x$ direction. The cyan arrow indicates the direction of motion of skyrmions.
(a) A current with $\theta=141^{\circ}$ is applied to drive skyrmions. The intrinsic skyrmion Hall angle $\theta_{\text{SkHE}}=-60^{\circ}$. Top-view snapshots at $t=300$, $680$, $900$, $1500$, $1930$, $3060$, and $4995$ ps are given.
(b) A current with $\theta=171^{\circ}$ is applied to drive skyrmions. The intrinsic skyrmion Hall angle $\theta_{\text{SkHE}}=-30^{\circ}$. Top-view snapshots at $t=160$, $320$, $430$, $600$, $795$, $905$, and $1995$ ps are given.
(c) A current with $\theta=231^{\circ}$ is applied to drive skyrmions. The intrinsic skyrmion Hall angle $\theta_{\text{SkHE}}=30^{\circ}$. Top-view snapshots at $t=80$, $240$, $450$, $600$, $700$, $770$, and $1000$ ps are given.
(d) A current with $\theta=261^{\circ}$ is applied to drive skyrmions. The intrinsic skyrmion Hall angle $\theta_{\text{SkHE}}=60^{\circ}$. Top-view snapshots at $t=100$, $350$, $600$, $800$, $1000$, $1200$, and $2000$ ps are given.
}
\label{FIG11}
\end{figure*}

\begin{figure*}[t]
\centerline{\includegraphics[width=0.98\textwidth]{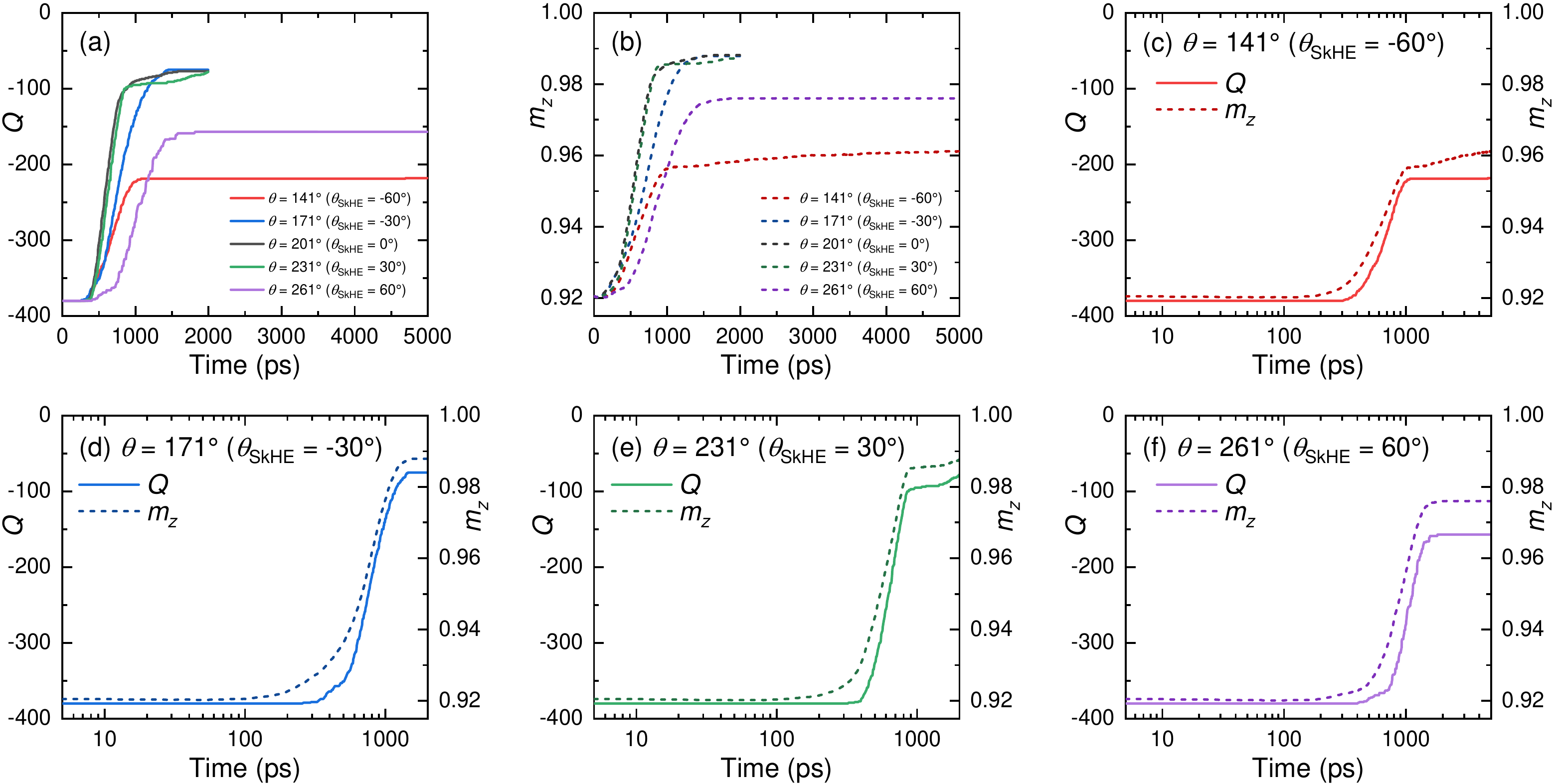}}
\caption{%
Time-dependent skyrmion number $Q$ and reduced out-of-plane magnetization $m_z$ for the systems driven by $j=0.5$ MA cm$^{-2}$ with different spin-polarization angle $\theta$. The apex of the funnel obstacle is pointing to the $-x$ direction.
(a) Time-dependent $Q$ for different $\theta$.
(b) Time-dependent $m_z$ for different $\theta$.
(c) Time-dependent $Q$ and $m_z$ for $\theta=141^{\circ}$, where the intrinsic skyrmion Hall angle $\theta_{\text{SkHE}}=-60^{\circ}$.
(d) Time-dependent $Q$ and $m_z$ for $\theta=171^{\circ}$, where $\theta_{\text{SkHE}}=-30^{\circ}$.
(e) Time-dependent $Q$ and $m_z$ for $\theta=231^{\circ}$, where $\theta_{\text{SkHE}}=30^{\circ}$.
(f) Time-dependent $Q$ and $m_z$ for $\theta=261^{\circ}$, where $\theta_{\text{SkHE}}=60^{\circ}$.
The simulation time is $5000$, $2000$, $2000$, and $5000$ ps, for the systems with $\theta=141^{\circ}$, $171^{\circ}$, $231^{\circ}$, and $261^{\circ}$, respectively.
}
\label{FIG12}
\end{figure*}

\section{Results and Discussion}
\label{se:Results}

\subsection{The skyrmion Hall angle}
\label{se:SkHE}

In order to facilitate the interaction between the funnel obstacle and skyrmions in our system, we first study the intrinsic skyrmion Hall angle of a single skyrmion driven by the damping-like torque $\boldsymbol{\tau}_{\text{d}}$.
The reason is that the direction of motion of skyrmions is determined by the intrinsic skyrmion Hall angle before encountering the obstacle or the edge of the ferromagnetic layer.
Hence we place a relaxed isolated skyrmion with $Q=-1$ at the center of a clean sample (i.e., at $x=1000$ nm and $y=500$ nm) and drive the skyrmion into motion by the damping-like torque with $\boldsymbol{p}$, where we define the spin-polarization angle between the spin-polarization direction $\boldsymbol{p}$ and the $+x$ direction as $\theta$. The value of $\theta$ ranges from $0^{\circ}$ to $360^{\circ}$ and can be controlled by changing the injection direction of a uniform electron flow $\boldsymbol{\hat{j}}_{e}$ in the heavy metal layer as $\boldsymbol{p}=\boldsymbol{\hat{j}}_{e}\times\boldsymbol{n}$ with $\boldsymbol{n}$ being the surface normal vector of the ferromagnet/heavy metal interface~\cite{Tomasello_SREP2014,Wanjun_SCIENCE2015,Sampaio_NN2013,Sinova_2015}.
The skyrmion number $Q$ is defined as $Q=\frac{1}{4\pi}\int\boldsymbol{m}\cdot(\frac{\partial\boldsymbol{m}}{\partial x}\times\frac{\partial\boldsymbol{m}}{\partial y})dxdy$~\cite{Nagaosa_NNANO2013,Zhang_JPCM2020,Gobel_PP2021} and could be calculated in a lattice-based approach in the \textsc{mumax$^3$} simulator~\cite{MuMax,LatticeQ}.

We numerically find that the skyrmion Hall angle $\theta_{\text{SkHE}}$ induced by a moderate current density $j=0.5$ MA cm$^{-2}$ at $\theta=0^{\circ}$ (i.e., $\boldsymbol{p}=+\hat{x}$) is equal to $158.97^{\circ}$, where we numerically define the skyrmion Hall angle as the angle between the skyrmion velocity vector and the $+x$ direction, i.e., $\theta_{\text{SkHE}}=\text{atan2}(v_y,v_x)$.
We note that the simulated skyrmion has a relaxed diameter of $\sim 15$ nm and shows no deformation during its motion driven by a small or moderate current density, justifying the rigidity of particle-like skyrmions considered in our system.
We then obtain the $\theta$-dependent intrinsic skyrmion Hall angle $\theta_{\text{SkHE}}$($\theta$) in our system according to the fact that the angle between the spin polarization direction $\boldsymbol{p}$ and the skyrmion velocity $v$ is a fixed value, as shown in Fig.~\ref{FIG1}.
The skyrmion motion is toward the right side of the ferromagnetic layer (i.e., $\theta_{\text{SkHE}}=-90^{\circ}-90^{\circ}$), when the spin-polarization angle is $\theta=112^{\circ}-291^{\circ}$.
We note that the intrinsic skyrmion Hall angle in the absence of pinning is independent of the driving current density provided that the skyrmion is not deformed by the driving force, which can also be understood from the Thiele equation~\cite{Thiele_PRL1973,Tomasello_SREP2014,Wang_PRB2019}, given as
\begin{equation}
\boldsymbol{G}\times\boldsymbol{v}-\alpha\boldsymbol{\D}\cdot\boldsymbol{v}-4\pi\boldsymbol{\B}\cdot\boldsymbol{j}_{e}=\boldsymbol{0},
\label{eq:TME-CPP}
\end{equation}
where $\boldsymbol{v}=(v_x,v_y)$ is the steady skyrmion velocity and $\boldsymbol{j}_{e}=(j_x,j_y)$ is the electron current that produces the damping-like torque.
$\boldsymbol{G}=(0,0,-4\pi Q)$ is the gyromagnetic coupling vector associated with the Magnus force term.
$\boldsymbol{\D}=4\pi\begin{pmatrix} \D_{xx} & \D_{xy} \\ \D_{yx} & \D_{yy} \end{pmatrix}$ is the dissipative tensor for the dissipative force term, where we have $\D_{xx}=\D_{yy}=\D$ and $\D_{xy}=\D_{yx}=0$ in our system.
$\boldsymbol{\B}=\frac{u}{j_{e}}\begin{pmatrix} -\I_{xy} & \I_{xx} \\ -\I_{yy} & \I_{yx} \end{pmatrix}$ is a term that quantifies the efficiency of the spin torque over the skyrmion, where we have $\I_{xx}=\I_{yy}=0$ and $\I_{xy}=-\I_{yx}=\I$ in our system.
Therefore, from Eq.~(\ref{eq:TME-CPP}) we find that~\cite{Wang_PRB2019}
\begin{equation}
v_x = \left|\frac{u}{j_{e}}\right|\I\frac{\alpha\D j_x+Qj_y}{Q^2+\alpha^2\D^2}, \quad
v_y = \left|\frac{u}{j_{e}}\right|\I\frac{\alpha\D j_y-Qj_x}{Q^2+\alpha^2\D^2}.
\label{eq:TME-CPP-vx-vy}
\end{equation}
The intrinsic skyrmion Hall angle could be straightforwardly defined as $\theta_{\text{SkHE}}=\arctan(v_y/v_x)$, so one could have
\begin{equation}
\theta_{\text{SkHE}}=\arctan(\frac{\alpha\D j_y-Qj_x}{\alpha\D j_x+Qj_y}).
\label{eq:TME-CPP-SkHE}
\end{equation}
As $\boldsymbol{p}=\boldsymbol{\hat{j}}_{e}\times\boldsymbol{n}$ and the angle between $\boldsymbol{p}$ and the $+x$ direction is $\theta$, we could have
$j_x=j_e\cos(\theta+90^{\circ})=-j_e\sin\theta$ and
$j_y=j_e\sin(\theta+90^{\circ})=j_e\cos\theta$.
Hence, the skyrmion Hall angle Eq.~(\ref{eq:TME-CPP-SkHE}) could be rewritten as a function of $\theta$, expressed as
\begin{equation}
\theta_{\text{SkHE}}=\arctan(\frac{\alpha\D\cos\theta+Q\sin\theta}{-\alpha\D\sin\theta+Q\cos\theta}).
\label{eq:TME-CPP-SkHE-Theta}
\end{equation}
From Eq.~(\ref{eq:TME-CPP-SkHE-Theta}) it can be seen that the skyrmion Hall angle is theoretically independent of the driving current density.
Also, one should have $\theta_{\text{SkHE}}=\arctan(\alpha\D/Q)$ when $\theta=0$.
As we numerically find that $\theta_{\text{SkHE}}=158.97^{\circ}$ at $\theta=0^{\circ}$, $\alpha=0.3$, and $Q=-1$, we obtain that $\D=1.28$ for the skyrmion studied in our system.
Therefore, Eq.~(\ref{eq:TME-CPP-SkHE-Theta}) also suggests that $\theta_{\text{SkHE}}\sim 0^{\circ}$ at $\theta=201^{\circ}$, i.e., the skyrmion moves toward the $+x$ direction when $\theta=201^{\circ}$.
In Sec.~\ref{se:Funnel}, we will first focus on the skyrmion dynamics driven by a current in the system with $\theta=201^{\circ}$, where a large number of skyrmions will move with $\theta_{\text{SkHE}}=0^{\circ}$ and interact with a funnel obstacle with an angle of incidence of zero.

\subsection{Skyrmions interacting with a funnel obstacle}
\label{se:Funnel}

We first study the interactions between a large number of particle-like skyrmions and a funnel obstacle in the ferromagnetic layer. Technically the funnel obstacle is a region locally modified to have an enhanced PMA $K_{\text{o}}$. The region with enhanced PMA is able to repel the skyrmions outside but closely around the region~\cite{Juge_NL2021,Ohara_NL2021}. We assume that $K_{\text{o}}/K=10$ in order to make sure that the skyrmions in our system cannot penetrate the boundary of the funnel obstacle.

Figure~\ref{FIG2} shows the initial spin configuration as well as the configuration of the funnel obstacle in the ferromagnetic layer.
The initial state is a triangular lattice of $380$ relaxed particle-like skyrmions with $Q=-1$ at the left side of the ferromagnetic layer (i.e., $x=0-1000$ nm and $y=0-1000$ nm).
The funnel obstacle is placed at the right side of the ferromagnetic layer with its apex pointing to the $+x$ or $-x$ direction. In this work, the apex angle is set to $60^{\circ}$ and the tip-to-tip width of the funnel obstacle is set to $\sim 578$ nm. The wall width of the funnel obstacle is set to $\sim 25$ nm.
We note that the periodic boundary condition applied in the $y$ direction can avoid the interactions between the skyrmions and the top and bottom edges of the ferromagnetic layer and, therefore, can ensure that most skyrmions can interact with the funnel obstacle effectively. By considering the periodic boundary condition in the $y$ directions, the skyrmions interacting with the funnel obstacle in the system may be treated as the interactions between the skyrmions and a one-dimensional array of funnel obstacles at certain conditions. From this point of view, the opening size between the adjacent tips of two funnel obstacles would be equal to $\sim 425$ nm in our system.

We first study the interactions between the particle-like skyrmions and the funnel obstacle with its apex pointing to the $+x$ direction [see Fig.~\ref{FIG2}(a)].
As discussed in Sec.~\ref{se:SkHE}, we apply a current with $\theta=201^{\circ}$ to ensure the current-driven motion of skyrmions toward the $+x$ direction, i.e., $\theta_{\text{SkHE}}=0^{\circ}$.
Namely, the angle of incidence of the particle-like skyrmions on the line connecting two funnel tips near $x=1000$ nm is almost zero.
In all simulations, the current is turned on at $t=0$ ps to drive skyrmions.

As shown in Fig.~\ref{FIG3}(a), we first apply a small current of $j=0.3$ MA cm$^{-2}$ to drive the skyrmions toward the funnel obstacle [see Video \blue{1} in the Supplemental Material (SM)~\cite{SM}].
A large part of the skyrmions is moving into the triangular region between the upper and lower funnel walls when the current is turned on at $t=0$ ps. A line of skyrmions close to the upper funnel wall moves faster along the wall and reaches the apex of the funnel obstacle at $t=145$ ps.
The increase of the skyrmion speed $v$ along the upper funnel wall is a result of the skyrmion-funnel repulsion, i.e., similar to the skyrmion-edge repulsion~\cite{Iwasaki_NL2014,Xichao_PRB2016B,Tomasello_SREP2014}. The repulsive force acting on the skyrmion leads to the skyrmion motion along the wall and generally toward the $-y$ direction (i.e., $v_y<0$).
When more skyrmions move into and accumulate in the triangular region between the upper and lower funnel walls, a clogging and compression effects of the skyrmions are found due to the repulsive skyrmion-funnel and skyrmion-skyrmion interactions [cf. snapshots at $t=200$ ps and $t=530$ ps in Fig.~\ref{FIG3}(a)].
The compression of skyrmions could lead to the decrease of skyrmion size and may result in the annihilation or collapse of skyrmions. Also, the decrease of skyrmion size will result in the increase of the reduced out-of-plane magnetization of the ferromagnetic layer because the skyrmion with $Q=-1$ in our system is stabilized in a background magnetization pointing to the $+z$ direction but has a core with magnetization pointing to the $-z$ direction.
In Figs.~\ref{FIG4}(a) and~\ref{FIG4}(b), we show the time-dependent skyrmion number $Q$ and reduced out-of-plane magnetization component $m_z$ of the system, respectively. It can be seen that $Q$ remains unchanged before $t=530$ ps, while $m_z$ increases with time before $t=530$ ps [see Fig.~\ref{FIG4}(c)], which means that the skyrmions in the ferromagnetic layer are compressed but not annihilated.
It is noteworthy that the compression of skyrmions in the triangular region between the upper and lower funnel walls also leads to the formation of a compressed triangular lattice of skyrmions, as shown in Fig.~\ref{FIG5}.

Many skyrmions in the triangular region between the upper and lower funnel walls, especially those near the funnel apex, are not moving during the compression, which may be referred to as a clogging effect of skyrmions, although many skyrmions could move toward the right edge of the ferromagnetic layer through the openings between the top/bottom ferromagnetic layer edge and the upper/lower funnel tip.
When the compression effect is over (i.e., no more skyrmions are driven into the triangular region), the skyrmions inside the triangular region between the upper and lower funnel walls could escape from the triangular region via moving slowly along the lower funnel wall, showing a negative $v_x$ and $v_y$.
Once the skyrmions are out of the triangular region, they will move toward the right edge of the ferromagnetic layer. Especially, the skyrmions moving along the lower funnel wall will show a U-turn at the lower funnel tip, as indicated by the cyan arrows in Fig.~\ref{FIG3}(a) at $t=800$ ps.

We also note that the skyrmions arrive at the right edge of the sample at $t\sim 530$ ps. As periodic and open boundary conditions are applied in the $y$ and $x$ directions, respectively, the skyrmions will accumulate near the right edge of the ferromagnetic layer and form a flow of skyrmions moving along the edge toward the $-y$ direction [see Fig.~\ref{FIG3}(a); $t=800-2300$ ps].
The motion of skyrmions along the right edge toward the $-y$ direction is a result of the skyrmion-edge interaction, where the skyrmion velocity in the $x$ direction is limited to zero (i.e., $v_x=0$) due to the skyrmion-edge repulsion~\cite{Iwasaki_NL2014,Xichao_PRB2016B,Tomasello_SREP2014}.
When more skyrmions that have escaped from the triangular region between the upper and lower funnel walls are driven to the right edge of the ferromagnetic layer at $t>800$ ps, a strong compression effect due to the repulsive skyrmion-edge and skyrmion-skyrmion interactions leads to the annihilation of many skyrmions near the right edge of the ferromagnetic layer, which can be seen from the time-dependent $Q$ and $m_z$ in Fig.~\ref{FIG4}(c).
Both $Q$ and $m_z$ increase with time in the system after $t\sim 800$ ps, indicating the fast compression-induced annihilation of skyrmions at the right edge of the ferromagnetic layer.

When most skyrmions in the triangular region between the upper and lower funnel walls have escaped, a line of several skyrmions moves along the lower funnel wall and gets out of the funnel obstacle from the lower funnel tip, as shown in Fig.~\ref{FIG3}(a) at $t=2300$ ps.
Once the last few skyrmions are out of the funnel obstacle, they form a beam of skyrmions moving toward the $+x$ direction, which finally get merged into the skyrmion flow moving along the right edge of the ferromagnetic layer.
We note that the stable number of skyrmions moving along the right edge of the ferromagnetic layer, i.e., the number of skyrmions that could remain in the system, is determined by the driving force, as shown in Fig.~\ref{FIG4}(a). Namely, a larger driving current $j$ will result in a stronger compression and thus, the annihilation of more skyrmions.

In Fig.~\ref{FIG3}(b), we show the skyrmion dynamics interacting with a funnel obstacle in the system driven by an increased current of $j=0.5$ MA cm$^{-2}$ (see SM Video \blue{2}~\cite{SM}). The funnel apex is pointing to the $+x$ direction.
The dynamic behaviors of the particle-like skyrmions interacting with a funnel obstacle driven by $j=0.5$ MA cm$^{-2}$ are quantitatively similar to that induced by a smaller current of $j=0.3$ MA cm$^{-2}$.
However, we note that an obvious difference is that the skyrmions in the ferromagnetic layer start to annihilate before they arrive at the right edge of the ferromagnetic layer at $t\sim 325$ ps, which can be seen from the time-dependent $Q$ and $m_z$ in Fig.~\ref{FIG4}(d).
The reason is that the increased driving force leads to stronger compression of skyrmions in the triangular region between the upper and lower funnel walls and thus further leads to the collapse of skyrmions near the upper and lower funnel walls.
The increased driving force also results in stronger compression and more significant annihilation of skyrmions near the right edge of the ferromagnetic layer.
For example, the snapshot at $t=1350$ ps in Fig.~\ref{FIG3}(b) shows that the dynamically stable skyrmion flow moving along the right edge of the ferromagnetic layer consists of three vertical lines of compactly arranged skyrmions, while it consists of five vertical lines of skyrmions in Fig.~\ref{FIG3}(a) at $t=2300$ ps.

In Figs.~\ref{FIG3}(c) and~\ref{FIG3}(d), we show the skyrmion dynamics interacting with a funnel obstacle in the system driven by further increased current densities of $j=0.7$ MA cm$^{-2}$ (see SM Video \blue{3}~\cite{SM}) and $j=0.9$ MA cm$^{-2}$ (see SM Video \blue{4}~\cite{SM}), respectively. The funnel apex is pointing to the $+x$ direction.
The corresponding time-dependent $Q$ and $m_z$ driven by $j=0.7$ MA cm$^{-2}$ and $j=0.9$ MA cm$^{-2}$ are given in Figs.~\ref{FIG4}(e) and~\ref{FIG4}(f), respectively.
The current-driven skyrmion dynamics interacting with a funnel obstacle are quantitatively similar to that induced by a smaller $j$, but the overall skyrmion speed is much higher at $j=0.7$ MA cm$^{-2}$ and $j=0.9$ MA cm$^{-2}$.
The large driving force also leads to the strong compression and annihilation of skyrmions, first in the triangular region between the upper and lower funnel walls and then in the region near the right edge of the ferromagnetic layer.
As the skyrmion speed is very fast at $j=0.9$ MA cm$^{-2}$, the initial compression and annihilation of skyrmions in the triangular region almost happen at the same time and, therefore, the time-dependent $Q$ and $m_z$ curves show similar trends in Fig.~\ref{FIG4}(f).
It is worth mentioning that a clogging effect of skyrmions could be found in the region between the lower funnel tip and the bottom edge of the ferromagnetic layer, where the skyrmions escaped from the triangular region between the upper and lower funnel walls and the skyrmions moving along the bottom edge will converge to form a flow of skyrmions in the form of a circular sector (see Fig.~\ref{FIG3}). Such a clogging effect of skyrmions could be reduced at larger $j$ or when almost all skyrmions in the triangular region between the upper and lower funnel walls have escaped.
It should also be noted that the interactions between the skyrmions and the funnel obstacle can be more generally regarded as the interactions between the skyrmions and a one-dimensional array of funnel obstacles because of the periodic boundary condition applied in the $y$ direction of the ferromagnetic layer.

We continue to investigate the interaction between the particle-like skyrmions and the funnel obstacle with its apex pointing to the $-x$ direction [see Fig.~\ref{FIG2}(b)].
In the same way, we apply a current with $\theta=201^{\circ}$ to ensure the current-driven motion of skyrmions toward the $+x$ direction, i.e., $\theta_{\text{SkHE}}=0^{\circ}$ (see Sec.~\ref{se:SkHE}).
Namely, the angle of incidence of the skyrmions on the line connecting two funnel tips near $x=1000$ nm is almost zero.

We first apply a small current of $j=0.3$ MA cm$^{-2}$ to drive the skyrmions toward the funnel obstacle (see SM Video \blue{5}~\cite{SM}).
When the skyrmions meet the funnel obstacle, they will be separated into two flows, as shown in Fig.~\ref{FIG6}(a) at $t=160$ ps. The skyrmions close to the lower funnel wall move faster and a line of skyrmions moving along the lower funnel wall first reaches the lower funnel tip at $t=160$ ps.
However, the skyrmions close to the upper funnel wall move slower and show a clogging effect near the upper funnel wall. Some skyrmions near the upper funnel tip move toward the $+x$ direction, while some skyrmions near the funnel apex first move toward the $-y$ direction and then move along the lower funnel wall toward the $+x$ direction [see Fig.~\ref{FIG6}(a) at $t=300$ ps].
The difference between the dynamics of skyrmions near the upper and lower funnel walls is due to the repulsive skyrmion-funnel interaction, similar to the skyrmion dynamics along the right edge of the ferromagnetic layer. Namely, the repulsive force from the funnel wall or ferromagnetic layer edge acting on the skyrmion leads to the motion of skyrmion along the funnel wall or ferromagnetic layer edge toward the $-y$ direction~\cite{Iwasaki_NL2014,Xichao_PRB2016B,Tomasello_SREP2014}.

As some skyrmions in the upper half of the ferromagnetic layer and close to the upper funnel wall will move to the lower half of the ferromagnetic layer when interacting with the funnel obstacle, the skyrmion flow moving toward the right ferromagnetic layer edge in the lower half of the ferromagnetic layer has more skyrmions and forms an obvious clogging effect around the lower tip of the funnel obstacle, as shown in Fig.~\ref{FIG6}(a) at $t=300-1100$ ps.
The clogging of skyrmions around the lower tip of the funnel obstacle leads to a skyrmion flow area in the shape of a circular sector.
Such a phenomenon could also be found in the system with a funnel obstacle with its apex pointing to the $+x$ direction (see Fig.~\ref{FIG3}).

On the other hand, the moving skyrmions near the upper and lower funnel walls are compactly arranged in a triangular lattice due to the compression effect. The compression effect could lead to a reduced spacing between neighboring skyrmions as well as a reduced size of skyrmions; however, the compression and clogging of skyrmions around the funnel obstacle do not result in the annihilation of skyrmions, which can also be found by comparing the time-dependent $Q$ and $m_z$ given in Figs.~\ref{FIG7}(a)-\ref{FIG7}(c).
It can be seen that $Q$ remains unchanged before $t\sim 700$ ps, while $m_z$ starts to increase from $t\sim 200$ ps when skyrmions meet the funnel obstacle, indicating the compression-induced decrease of the skyrmion size.
The skyrmions are not annihilated under compression around the funnel obstacle because they are driven into motion along the funnel wall, which is different to the situation in Fig.~\ref{FIG3}, where skyrmions may be captured by the funnel obstacle and annihilated near the funnel apex.

The skyrmions in the system first arrive at the right edge of the ferromagnetic layer at $t\sim 400$ ps, and the skyrmions near the right edge will move toward the $-y$ direction to form a flow of compactly arranged skyrmions [see Fig.~\ref{FIG6}(a) at $t=560$ ps]. Such a phenomenon is caused by the driving force and the repulsive skyrmion-edge and skyrmion-skyrmion interactions.
The compression effect near the right edge of the ferromagnetic layer may result in the annihilation of skyrmions from the right edge, until the system reaches a dynamically equilibrium state where a certain number of skyrmions could move along the right edge [see Fig.~\ref{FIG7}(a)].

As shown in Fig.~\ref{FIG6}(a) at $t=560$, $800$, and $1100$ ps, the driving force could push the skyrmions onto the outer surface of the upper funnel wall, where the skyrmions near the funnel wall and close to the upper funnel tip move toward the $+x$ direction, while the skyrmions near the funnel apex first move toward the $-y$ direction and are then merged into the flow of skyrmions moving along the lower funnel wall toward the $+x$ direction.
When there are only a single line of skyrmions attached to the upper funnel wall [see Fig.~\ref{FIG6}(a) at $t=1400$ ps], the skyrmions move along the outer surface of the funnel obstacle and form a beam of skyrmions toward the right edge of the ferromagnetic layer.

We also study the skyrmion dynamics interacting with a funnel obstacle in the systems driven by larger current densities. The results induced by $j=0.5$, $0.7$, and $0.9$ MA cm$^{-2}$ are given in Figs.~\ref{FIG6}(b),~\ref{FIG6}(c), and~\ref{FIG6}(d), respectively (see SM Videos \blue{6}-\blue{8}~\cite{SM}). Besides, the corresponding time-dependent $Q$ and $m_z$ are given in Figs.~\ref{FIG7}(d),~\ref{FIG7}(e), and~\ref{FIG7}(f), respectively.
When the driving current is increased, the skyrmions move faster toward the $+x$ direction, while the dynamic behaviors of the skyrmions interacting with the funnel obstacle are qualitatively similar to that driven by a small $j=0.3$ MA cm$^{-2}$ [see Fig.~\ref{FIG6}(a)].
However, the compression of skyrmions around the outer surface of the funnel obstacle as well as the compression of skyrmions near the right edge of the ferromagnetic layer are more significant at a larger $j$.
The faster skyrmion speed and stronger compression of skyrmions due to the skyrmion-funnel and skyrmion-skyrmion repulsions lead to the annihilation of more skyrmions, which can be seen from the time-dependent curves of $Q$ and $m_z$ given in Fig.~\ref{FIG7}.
Indeed, the arrangement of skyrmions around the funnel obstacle will be more compact at a larger $j$.
For example, Fig.~\ref{FIG8}(a) shows an enlarged view of the system driven by $j=0.5$ MA cm$^{-2}$ at $t=180$ ps [see Fig.~\ref{FIG6}(b)].
In Fig.~\ref{FIG8}(b), we also show a Voronoi cell construction calculated based on the state given in Fig.~\ref{FIG8}(a).
It can be seen that the density of skyrmions near the outer surface of the funnel obstacle is obviously higher than that of the incoming skyrmions at the left side of the ferromagnetic layer.

\subsection{The effect of the angle of incidence of incoming skyrmions}
\label{se:Funnel-Angle}

We have studied the particle-like skyrmion dynamics interacting with a funnel obstacle, where the angle of incidence of incoming skyrmions is adjusted to be zero (i.e., the intrinsic skyrmion Hall angle $\theta_{\text{SkHE}}\sim 0^{\circ}$) by setting the spin-polarization angle as $\theta=201^{\circ}$.
In this section, we continue to investigate the skyrmion-funnel interaction for the systems where the angle of incidence of skyrmions is nonzero.
Namely, the intrinsic skyrmion Hall angle is controlled to be in the range of $\theta_{\text{SkHE}}=-90^{\circ}-90^{\circ}$, which could be realized by setting the spin-polarization angle to be $\theta=112^{\circ}-291^{\circ}$ as discussed in Sec.~\ref{se:SkHE}.

We first focus on the system with a funnel obstacle with its apex pointing to the $+x$ direction [see Fig.~\ref{FIG2}(a)], where we apply a moderate driving force of $j=0.5$ MA cm$^{-2}$ to drive the skyrmions toward the funnel obstacle.
As discussed in Sec.~\ref{se:Funnel}, some skyrmions may be annihilated in the system driven by $j=0.5$ MA cm$^{-2}$ due to the skyrmion-funnel interaction, but the computational simulation time could be reduced due to faster skyrmion speed.

As shown in Fig.~\ref{FIG9}(a), by setting the spin-polarization angle $\theta=141^{\circ}$, the intrinsic skyrmion Hall angle is adjusted to be $\theta_{\text{SkHE}}=-60^{\circ}$.
The skyrmions interacting with the funnel obstacle generate two flows of skyrmions (see SM Video \blue{9}~\cite{SM}). A flow of skyrmions guided by the upper funnel wall move toward the right edge of the ferromagnetic layer [see Fig.~\ref{FIG9}(a) at $t=680$ ps].
The other flow guided by the lower funnel wall move toward the bottom edge of the ferromagnetic layer, which form a steady flow in the $y$ direction due to the periodic boundary condition in the $y$ direction.
The skyrmions near the right edge of the ferromagnetic layer move along the right edge toward the $-y$ direction due to the skyrmion-edge and skyrmion-skyrmion repulsions, as discussed in Sec.~\ref{se:Funnel}. Some skyrmions are annihilated due to the compression near the right edge. The annihilation of skyrmions in the system can be seen from the time-dependent $Q$ and $m_z$ given in Figs.~\ref{FIG10}(a)-\ref{FIG10}(c).

When most skyrmions are moved to the right edge of the ferromagnetic layer, some skyrmions remain in the flow in the $y$ direction, which are guided by the upper and lower funnel walls [see Fig.~\ref{FIG9}(a) at $t=4990$ ps].
The skyrmions guided by the upper funnel wall have the trend to move toward the upper funnel tip, which is a result of the skyrmion-funnel repulsion and the driving force.
The skyrmions guided by the lower funnel wall move toward the lower funnel tip.
Although some skyrmions near the funnel apex may move toward the right edge of the ferromagnetic layer once they are detached from the upper funnel wall due to the skyrmion-skyrmion compression, a certain number of skyrmions could remain in the steady flow in the $y$ direction guided by the funnel walls.
Considering the periodic boundary condition in the $y$ direction, such a phenomenon can be regarded as the capturing of skyrmions by a one-dimensional array of funnel obstacles due to the repulsive skyrmion-funnel interaction.

In Fig.~\ref{FIG9}(b), by setting the spin-polarization angle $\theta=171^{\circ}$, the intrinsic skyrmion Hall angle is adjusted to be $\theta_{\text{SkHE}}=-30^{\circ}$ (see SM Video \blue{10}~\cite{SM}).
The skyrmions driven by the current move toward the right edge of the ferromagnetic layer, and most skyrmions are annihilated due to the compression of skyrmions near the right edge, as can be seen in Fig.~\ref{FIG10}(d).
The results are similar to the system with $\theta=201^{\circ}$ (i.e., $\theta_{\text{SkHE}}=0^{\circ}$) [see Fig.~\ref{FIG3}(b)].

In Fig.~\ref{FIG9}(c), by setting the spin-polarization angle $\theta=231^{\circ}$, the intrinsic skyrmion Hall angle is adjusted to be $\theta_{\text{SkHE}}=30^{\circ}$ (see SM Video \blue{11}~\cite{SM}).
The skyrmions driven by the current move toward the right edge of the ferromagnetic layer, while a large number of skyrmions are collected by the triangular region between the upper and lower funnel walls.
The compression of skyrmions inside the triangular region between the upper and lower funnel walls results in the annihilation of some skyrmions [see Fig.~\ref{FIG9}(c) at $t=330$ ps], however, the compression effect is reduced when no more skyrmions are driven into the triangular region [see Fig.~\ref{FIG9}(c) at $t=470$ ps].
The annihilation of skyrmions inside the triangular region between the upper and lower funnel walls leads to an early decrease in the absolute value of $Q$, as shown in Fig.~\ref{FIG10}(e).
Note that a few skyrmions move along the lower funnel wall toward the right edge of the ferromagnetic layer at $t=1500-1995$ ps, which are skyrmions overflowed from the triangular region between the upper and lower funnel walls.

In Fig.~\ref{FIG9}(d), by setting the spin-polarization angle $\theta=261^{\circ}$, the intrinsic skyrmion Hall angle is adjusted to be $\theta_{\text{SkHE}}=60^{\circ}$ (see SM Video \blue{12}~\cite{SM}).
Many skyrmions are driven into the triangular region between the upper and lower funnel walls, which leads to obvious compression and annihilation of skyrmions. The annihilation of skyrmions can be seen from the time-dependent $Q$ and $m_z$ in Fig.~\ref{FIG10}(f).
Meanwhile, a flow of skyrmions moving toward the right edge of the ferromagnetic layer is formed and guided by the lower funnel wall until $t=1030$ ps.
As shown in Fig.~\ref{FIG9}(d) at $t=4995$ ps, a group of $44$ skyrmions are captured by the funnel obstacle ultimately, which forms a compressed triangular lattice inside the triangular region between the upper and lower funnel walls.
It is worth mentioning that such a capturing of skyrmions by the funnel obstacle is different from that shown in Fig.~\ref{FIG9}(a).
Namely, the skyrmions captured by the funnel obstacle are dynamic and not compressed at $\theta=141^{\circ}$ in Fig.~\ref{FIG9}(a), while they are static and compressed at $\theta=261^{\circ}$ in Fig.~\ref{FIG9}(d).
The capturing of static skyrmions inside the triangular region between the upper and lower funnel walls can be done by either a single isolated funnel obstacle or a one-dimensional array of funnel obstacles with reasonable funnel opening size.

Lastly, we focus on the system with a funnel obstacle with its apex pointing to the $-x$ direction [see Fig.~\ref{FIG2}(b)], where we apply a moderate driving force of $j=0.5$ MA cm$^{-2}$ to drive the skyrmions toward the funnel obstacle.
As shown in Fig.~\ref{FIG11}(a), by setting the spin-polarization angle $\theta=141^{\circ}$, the intrinsic skyrmion Hall angle is adjusted to be $\theta_{\text{SkHE}}=-60^{\circ}$ (see SM Video \blue{13}~\cite{SM}).
In such a configuration, the funnel obstacle is also able to capture a flow of skyrmions moving in the $y$ direction, which is mainly guided by the upper funnel wall, as shown in Fig.~\ref{FIG11}(a) at $t=4995$ ps.
However, when the spin-polarization angle is set to $\theta=171^{\circ}$, $231^{\circ}$, and $261^{\circ}$, the funnel obstacle cannot capture the skyrmions, and most skyrmions are annihilated due to the compression near the right edge of the ferromagnetic layer, as shown in Figs.~\ref{FIG11}(b)-\ref{FIG11}(d), respectively (see SM Videos \blue{14}-\blue{16}~\cite{SM}).
The annihilation of skyrmions in the system due to the compression effect can be seen from the time-dependent $Q$ and $m_z$ curves given in Fig.~\ref{FIG12}.

\section{Conclusion}
\label{se:Conclusion}

In conclusion, we have studied the dynamics of a larger number of particle-like skyrmions interacting with a funnel obstacle in a ferromagnetic layer with periodic boundary condition employed in the $y$ direction.
It is found that the interactions between the skyrmions and the funnel obstacle depend on several controllable factors, including the driving current density $j$, the spin-polarization angle $\theta$ (i.e., the intrinsic skyrmion Hall angle $\theta_{\text{SkHE}}$), and the orientation of the funnel obstacle.
For the system with $\theta=201^{\circ}$ (i.e., $\theta_{\text{SkHE}}=0^{\circ}$), we find that the interactions between the skyrmions and the funnel obstacle may result in the compression and annihilation of skyrmions near the funnel walls. The funnel walls can guide the motion of skyrmions and may lead to the clogging of skyrmions in the system.
In particular, due to the repulsive skyrmion-funnel interaction, the funnel obstacle can separate the incoming skyrmions with an angle of incidence of zero into two asymmetric flows, and form a single beam of skyrmions for a certain duration of time.
For the system with $\theta_{\text{SkHE}}\neq 0^{\circ}$, where the angle of incidence of the skyrmions is nonzero, a certain number of skyrmions can be effectively captured by the funnel obstacle in either a dynamic or static way, which can be controlled by adjusting the value of the spin-polarization angle $\theta$. The capturing of a dynamic flow of skyrmions is realized by the guided motion of skyrmions along the funnel walls, while the capturing of static and compressed skyrmions is realized by the repulsive skyrmion-funnel interaction inside the triangular region between the upper and lower funnel walls.
We note that the amount of skyrmions captured by the funnel obstacle could be controlled, in principle, by changing the driving current density, the spin-polarization angle, and the funnel geometry.
We also point out that the dynamic behaviors of skyrmions interacting with a funnel substrate have been studied in Refs.~\onlinecite{Souza_PRB2021,Souza_2022} using the particle model~\cite{Lin_PRB2013}, while our investigation was made by using the micromagnetic simulation, where the spin dynamics is controlled by the LLG equation [i.e., Eq.~(\ref{eq:LLGS-CPP})]. Under the framework of the micromagnetics, we are able to uncover new physical phenomena, such as the compression-induced annihilation of skyrmions, which are not revealed by the particle model~\cite{Lin_PRB2013}. Besides, the skyrmion-skyrmion and skyrmion-funnel interactions in the micromagnetic simulations are results of both the short-range and long-range interactions, including the effects of the ferromagnetic exchange interaction, interfacial DMI, PMA, and demagnetization, which are also subject to the deformation of skyrmions that is not considered in the particle model. The deformation of a skyrmion before its annihilation is a key dynamic feature for the skyrmion under compression near the funnel obstacle, which is important for understanding the dynamic physics of skyrmions interacting with obstacles in real magnetic materials, where skyrmions are not absolutely rigid under compression. However, a promising advantage of the particle model is that it can consider the complex dynamics of a massive number of particle-like skyrmions interacting with the obstacles~\cite{Reichhardt_2021}, which could be an impossible task for the micromagnetic simulation due to the limited computational power. We remark that a combination of the micromagnetic simulation and particle model may be able to provide a comprehensive view of the complex dynamics of large-scale skyrmion systems.
Our results could be useful for the understanding of particle-like skyrmions interacting with artificial obstacles in ferromagnetic films and may open a new way for the manipulation of particle-like skyrmion flow or beam. Our results also provide a way to capture skyrmions driven by dc currents.

\begin{acknowledgments}
X.Z. was an International Research Fellow of the Japan Society for the Promotion of Science (JSPS).
X.Z. was supported by JSPS KAKENHI (Grant No. JP20F20363).
J.X. was a JSPS International Research Fellow.
J.X. was supported by JSPS KAKENHI (Grant No. JP22F22061).
X.L. acknowledges support by the Grants-in-Aid for Scientific Research from JSPS KAKENHI (Grants No. JP20F20363, No. JP21H01364, No. JP21K18872, and No. JP22F22061).
\end{acknowledgments}




\end{document}